\documentclass[12pt]{article}
\usepackage{epsfig}
\newcommand{\beq}{\begin{equation}}
\newcommand{\eeq}{\end{equation}}
\newcommand{\beqn}{\begin{eqnarray}}
\newcommand{\eeqn}{\end{eqnarray}}
\newcommand{\eq}[1]{(\ref{#1})}

\newcommand{\dual}[1]{{}^{*}{#1}}

\newcommand{\dd}{{\mathrm d}}

\newcommand{\diff}{\partial}
\newcommand{\cC}{{\cal C}}
\newcommand{\cZ}{{\cal Z}}
\newcommand{\cD}{{\cal D}}

\newcommand{\bh}{{\bf H}}

\newcommand{\NPB}[3]{{\it Nucl. Phys. }{\bf B#1} (#2) #3}

\newcommand{\PLB}[3]{{\it Phys. Lett. }{\bf B#1} (#2) #3}
\newcommand{\PRL}[3]{{\it Phys. Rev. Lett. }{\bf #1} (#2) #3}

\newcommand{\PRD}[3]{{\it Phys. Rev. }{\bf D#1} (#2) #3}

\begin{document}
\date{}
\title{MONOPOLES AND CONFINING STRINGS IN QCD
\vskip-40mm
\rightline{\small MPI-PhT/2001-02}
\vskip 40mm
}

\author{
M.N. Chernodub$^{\rm a}$,
F.V.~Gubarev$^{\rm a,b}$,
M.I. Polikarpov$^{\rm a}$, 
V.I.~Zakharov$^{\rm b}$ \\
\\
$^{\rm a}$ {\small\it Institute of Theoretical and  Experimental Physics,}\\
{\small\it B.Cheremushkinskaya 25, Moscow, 117259, Russia}\\
$^{\rm b}$ {\small\it Max-Planck Institut f\"ur Physik,
F\"ohringer Ring 6, 80805 M\"unchen, Germany}
}

\maketitle

\begin{abstract}
{We review the recent results in the physics
of the magnetic monopoles in gluodynamics and a dual formulation of
non-Abelian theories, relevant to the physics of the confinement. It
occurs that the dual gluon is a U(1) gauge boson,  despite of the fact that
usual gluons are non-Abelian. The effective infrared Lagrangian for
gluodynamics is suggested which leads to the Casimir scaling of the
string tension for quarks in various representations. We also show
that the results of the calculations in
lattice gauge theories confirm our theoretical predictions.}
\end{abstract}

%=============================================================================

\section{Introduction}

Magnetic monopoles were introduced theoretically by Dirac  about 70 years
ago~\cite{dirac}. By definition, the radial magnetic field of the monopole
is similar to the electric field of a point-like charge:

\beq\label{monopole}
\vec{\bh}^{Coul} =  \frac{Q_m\vec{\bf r}}{4\pi r^3} \, ,
\eeq
where $Q_m$ is the magnetic charge. However, the field (\ref{monopole}) by
itself is inconsistent with the Maxwell equation $div{\bf H}=0$. To ensure
the conservation of the magnetic flux Dirac postulated, therefore, that the
magnetic field is transported to the monopole from infinity through a
string:

\beq \vec{\bh}~=~\vec{\bh}^{Coul}+\vec{\bh}^{Str}\, ,\,\,
\vec{\bh}^{Str} = - Q \hat{{\bf z}}\delta(x)\delta(y)\theta(z) ,
\eeq
where the Dirac string is directed along the negative {\it z} axis and $\hat{{\bf
z}}$ is the unit vector.  The corresponding monopole gauge potential has a
simple form in the spherical coordinates:

\beq\label{dirac}
A_r~=~A_{\theta}~=0, ~~~A_{\phi}~=~\frac{Q_m}{4\pi}{(1 + \cos \theta)\over
r\sin\theta}.
\eeq

Thus, monopoles cannot be introduced without a string attached to it and, at
first sight, any similarity between the electric and magnetic charges is
lost. The only way to nevertheless maintain the similarity between the two
types of charges is to ensure that the Dirac string is in fact unobservable.
Thus, Dirac derived a few constraints on the theory which should be
satisfied to make the monopoles effectively particle-like, not string-like
objects. Theoretically, the Dirac monopole and the Dirac strings are
fascinating subjects, for review see, e.g.,~\cite{coleman,blagoevic}. Still,
they look exotic objects with infinite energy and have never been observed.

The saga of the monopoles took a new turn with formulation of the (dual)
superconductor model of confinement~\cite{dualmodel}. It is well known that
the color states are not observable as free particles since  the
potential between two static quarks grows linearly with the distance $r$ at
large $r$ :

$$ \lim_{r\to \infty}{V_{q\bar{q}}}(r)~=~\sigma \cdot r~~$$
Note that the potential $V_{q\bar{q}}$ is defined in a gauge invariant
way in terms of the Wilson loop $W_{C}$ (for a pedagogical discussion and
further references see, e.g., Ref.~\cite{brambilla}). Namely,

\beq\label{vqq}
V_{q\bar{q}}~=~- \lim_{T\to\infty}{{1\over T}\log <W_{\Gamma_0}>},
\eeq
where
\beq\label{wilsonloop}
W_C~=~Tr P~exp \left({i\oint_C\hat{A}_{\mu}dx_{\mu}}\right)
\eeq
and $\hat{A}_{\mu}$ is the non-Abelian gauge potential while $\Gamma_0$ is a
rectangular contour $r\times T$.

Searching for an analogy to this spectacular phenomenon of confinement, one
 can notice that if magnetic monopoles, as test particles would be placed
into a superconductor the heavy monopole potential would grow linearly  as
well.  Indeed the magnetic field cannot penetrate the superconductor and
would stream into an Abrikosov vortex~\cite{abrikosov}.  The superconductor
model of confinement (for review see, e.g.,~\cite{baker,brambilla}) assumes
that the color electric field of the heavy quarks is organized into a
similar tube-like structure in the vacuum state of the gluodynamics. For the
model to realize, one needs condensation of the magnetic monopoles in the
QCD vacuum.

Although the model looks very appealing and, in a way, no alternative to it
has ever been found, it poses quite a few questions which seem very
difficult to answer.  First of all, we consider now a non-Abelian theory
while the (dual) superconductor model of the confinement copies the
electrodynamics, that is a $U(1)$ theory.  Moreover, there are no stable
classical monopole solutions to the pure Yang-Mills equations with a finite
energy~\cite{bn} and the condensation of the monopoles is difficult to
visualize.

The scope of this review does not allow even to mention the milestones on
the road to the resolution of the puzzles brought by the model
\cite{dualmodel}.  To make the story very short, the monopoles were
copiously observed and the dual superconductor model of confinement has been
confirmed, for review see, e.g.,~\cite{reviews}. However, there was a price
to be paid. Just to mention a few points:  the lattice regularization and
numerical methods seem crucial, as well as the choice of a particular gauge.

The need for the lattice regularization in the ultraviolet might look most
surprising since the motivation to introduce the monopoles was to explain
the confinement, i.e.  the basic feature of the physics in  the infrared.
However, there are good reasons for this. Indeed, the Dirac strings attached
to the monopoles are infinitely thin in the continuum limit. Therefore, the
Dirac constraints on the strings can be checked only within a particular
ultraviolet regularization.  Moreover, the gauge potentials associated with
the monopoles are singular both at the string and at small $r$, see, e.g.,
Eq.~(\ref{dirac}). In other words the monopoles, also in  non-Abelian theories
represent topological defects and the potentials are to be regularized in
the ultraviolet.

Moreover, the use of the numerical simulations is not only due to the
complexity of the vacuum state but, even more so, due to the lack of
understanding of a single monopole configuration.  Indeed, classically only
the monopoles with the minimal magnetic charge (in the units of the Dirac
quantization condition) are stable and have an infinite action~\cite{bn}.
The ``lattice monopoles'' observed in the numerical simulations would
correspond to unstable classical filed configurations and cannot be,
therefore, understood quasiclassically. In this respect the lattice
monopoles in non-Abelian theories are very different from instantons in case
of the same non-Abelian theories or Dirac monopoles in the compact $U(1)$
\cite{Polyakov-compact-U1}.

The use of the specific gauges also appears as a precondition for the
success of the dual superconductor model. Partly, this can be understood
theoretically.  Indeed, it was realized long time ago~\cite{thooft190} that
the monopoles can emerge as effective dynamical degrees of freedom only if
one fixes gauge up to a remaining $U(1)$ symmetry.  However, the numerical
simulations indicate strongly that various ways of fixing the remaining
$U(1)$ are not equivalent to each other at all as far as the phenomenology
is concerned.  The so called Maximal Abelian gauge turns to be most
successful~\cite{Abelian-Dominance,reviews}.

Schematically, the idea of the monopole condensation was formulated to
explain the basic feature of the  QCD as we know it in the real world, that
is confinement.  The idea was brought to the form suitable for the lattice
simulations and  dramatically confirmed through such simulations.  However,
the monopoles observed on the lattice look quite specific just for the
lattice formulation and we need now to address anew the continuum theory to
incorporate the lessons learnt through the simulations. One can say that the
pendulum is swinging now in the opposite direction:  from the lattice
formulation towards the continuum theory.  Broadly speaking, this review is
summarizing the recent developments in this direction.  While the review is
based mostly on the original papers in Ref~\cite{main}, it is worth
emphasizing that there are other papers devoted to the same or related
topics, see, e.g.,~\cite{ichie,kondo,marchetti}. Thus, it appears quite a
common trend and we believe that work in the direction of incorporating
lessons from the lattice studies into the continuum theory would continue in
the future.

More specifically, we will concentrate on  the question how to bridge the
$U(1)$ nature of the monopoles with the observation that they play a key
role in the dynamics of non-Abelian theories.  This double-face nature of
the monopoles is revealed most straightforwardly through study of the heavy
monopole potential $V_{m\bar{m}}(r)$, or the so called 't~Hooft loop
\cite{loop}.  The 't~Hooft loop allows to introduce a pair of (infinitely
heavy) external monopoles, like the Wilson loop (\ref{wilsonloop}) allows to
introduce a pair of external (infinitely heavy) quarks.  It is worth
emphasizing already at this early point that the monopoles introduced via
the 't~Hooft loop have the minimal magnetic charge $Q_m=1$ in the units of
the Dirac quantization condition while the lattice monopoles living in the
vacuum have $Q_m=2$.  In particular, the Dirac string corresponding to the
$Q_m=1$ monopoles is not visible to the gluons but would be visible to
quarks while the Dirac string attached to the $Q_m=2$ monopoles is not
visible both to particles in the adjoint and fundamental representations of
the color group.

The 't~Hooft loop operator is most easily formulated on the lattice
\cite{susskind}. The monopoles are understood then as the end-points of the
Dirac strings which in turn are defined as piercing negative plaquettes (see
Section~\ref{Hlooplat} for a detailed discussion). In the language of the
continuum theory it would be natural to define the 't~Hooft loop in terms of
the {\it dual gluon field} $B_{\mu}$:

\beq\label{hooftloop}
H_{C}~\equiv~\exp\left({i \,\frac{2\pi}{g}\,\int_{C} {B_{\mu}dx_{\mu}}}\right)~~,
\eeq
Then the heavy monopole potential is defined similarly to (\ref{vqq}):
\beq\label{vmm}
V_{m\bar{m}}~=~- \lim_{T\to\infty}{{1\over T}\log <H_{\Gamma_0}>}.
\eeq
The central point is of course, how to relate the ``dual gluon'' to the
fields entering the standard Lagrangian of the non-Abelian gauge theory.
And this is one of the central issues which we will try to clarify here
following~\cite{main}.

Generally, a dual gauge boson is understood as a field interacting with the
magnetic current $j_m$:

\beq
L_{int}~=~Q_mB\cdot j_m~~.
\eeq
One of our basic points is that the field $B$ should be treated as
fundamental in gluodynamics since the external monopoles introduced via the
't~Hooft loop are point like in the continuum limit.  A well known example
of introduction of a dual gauge boson is the Zwanziger Lagrangian describing
electrodynamics with both electric  and magnetic point like charges
\cite{zwanziger}.  Formally one introduces two vector fields, the
``standard'' photon $A_{\mu}$ and the dual photon $B_{\mu}$, so that
$L_{int}=Q_eA\cdot j_e+Q_mB\cdot j_m$.  However, the number of the degrees
of freedom is not changed since there is a constraint that the field
strength tensor constructed on the potential $A$ coincides with the dual
field strength tensor constructed on the potential $B_{\mu}$.  More
precisely:

\beq\label{zwanziger}
m_{\mu}F_{\mu\nu}(B)~=~m_{\mu}\dual{F_{\mu\nu}(A)}~~
\eeq
where $m_{\mu}$ is an arbitrary space-like vector. The choice of the vector
$m_{\mu}$ is a kind of new gauge freedom. Physically, one can visualize
$m_{\mu}$ as the vector along which all Dirac strings are directed.

In case of the gluodynamics, the dual gluon is still a $U(1)$ gauge boson
despite of the fact the direct gluons $A^a$ are in an adjoint representation
\cite{main}.  For simplicity we concentrate on the $SU(2)$ case. Then the
generalization of (\ref{zwanziger}) looks as:

\beq\label{zwanziger1}
m_{\mu}F_{\mu\nu}(B)~=~m_{\mu}\dual{(n^aF^a_{\mu\nu}(A))}
\eeq
where $F_{\mu\nu}^a$ is the non-Abelian field strength tensor and $n^a$ is
an arbitrary vector in the color space. Again, the choice of $n^a$ is a
matter of gauge fixing and the physical results do not depend on this
choice.  One can visualize the vector $n^a$ as the direction in the color
space of the magnetic fields transported along the Dirac strings attached to
the monopoles.

One can derive a Zwanziger-type Lagrangian which ensures the validity of the
constraint (\ref{zwanziger1}), see Section~\ref{dualgluon}. This
formulation allows then to make predictions for the heavy monopole potential
(\ref{vmm}). In particular, at short distances the potential can be derived
from the first principles:

\beq\label{vmm1}
V_{m \bar{m}}(R)~=~- {\bf 1} \cdot \frac{Q^2_m}{4\pi R}\, ,\,\,\,
Q_m\cdot g = 2\pi
\eeq
where $g$ is the standard coupling entering the non-Abelian Lagrangian.
To appreciate the meaning of (\ref{vmm1}) it is instructive to compare it
with the standard prediction for the heavy quark potential:
\beq\label{vqq1}
V_{q \bar{q}}(R)~=~- {\bf \frac34} \cdot \frac{g^2}{4\pi R}.
\eeq
Note that the overall coefficient in front of the potential, that is $3/4$,
reflects the non-Abelian nature of the quarks which belong to a fundamental
representation. There is no such factor in case of the monopole potential
(\ref{vmm1}). Thus, monopoles in $SU(2)$ gluodynamics are Abelian!

On the other hand, the non-Abelian nature of the theory gets manifested in
the running of the coupling. In the both cases (\ref{vqq1}) and (\ref{vmm1})
the quantum corrections result in $g^2\to g^2(R)$. Although the final result
is similar, the derivation is actually different. Namely, in case of the
radiative corrections to (\ref{vqq1}) one deals of course with a usual field
theory. In case of (\ref{vmm1}), performing the regularization procedure one
should address anew the role of the Dirac string attached to the heavy
monopoles and check that this procedure is respecting the Dirac constraints.
This is one of most subtle and exciting points and we are going to discuss
it in details in Sections~\ref{monoprenorm}, \ref{nabmonrenorm}. It is worth
emphasizing that there appearing first direct measurements of the
$V_{m\bar{m}}$ on the lattice~\cite{hoelbling} so that the prediction
(\ref{vmm1}) can be checked through measurements~\cite{main}.

The Abelian nature of the monopoles poses another serious problem which is
the theoretical understanding of the so called {\it Casimir scaling}
\cite{greensite}. First, let us remind the reader what the Casimir scaling
is.  One can measure the heavy quark potential not only for the sources in
the fundamental representation, as we discussed so far, but in any other
representation as well. At very large distances the potentials are expected
to depend crucially on the representation. For example, if the quarks have
isospin $T=1$ then at large distances the confining string connecting the
quarks should break into two mesons and the potential, respectively, is
expected to flatten out. However at all the distances available for the
lattice measurements so far an approximate equation for $V_{\bar{q}q}$
holds~\cite{bali}, \cite{shevchenko}:

\beq\label{casimir}
V_T(R)~\approx~ - T(T+1)\;{\alpha_s\over \pi~R}~+~T(T+1)\;
\sigma\,R \, ,
\eeq
where the string tension $\sigma$ does not depend on the isospin $T$.  The
Casimir scaling, at least at first sight, makes Abelian dominance very
questionable. Indeed, in, say, T=1 representation one quark is neutral with
respect to any singled out $U(1)$ subgroup of the $SU(2)$ and this quark
should have escaped any interaction, in bald violation of (\ref{casimir}).

We will argue~\cite{main} that in the approach where the choice of the
$U(1)$ subgroup is a matter of the gauge fixing, see above, the problem of
the Casimir scaling can be settled.  Within this approach the heavy quark
potential is:

\beq
V_T(R)~\approx~ - T(T+1)\;{\alpha_s\over \pi~R}~+~T(T+1)\;
\sigma_T(m_H,m_V)\,R \, ,
\eeq
where $m_H$, $m_V$ are the parameters of the dual model. The Casimir
scaling corresponds then to a particular limit, $m_H/m_V\gg 1$. In
Section~\ref{Casimir} we will discuss this problem in detail, along with
other related phenomenological issues.

In Section~\ref{Lattice} we show that the results of the numerical
calculations in lattice gauge theories confirm the dual superconductor model
of the gluodynamic vacuum. Moreover the recent study of the confining string
structure and monopole structure support the theoretical ideas given in this
review.

The review can be subdivided into two parts: Sections~\ref{monopoles} --
\ref{latticemonopoles} contain the introductory remarks and main
definitions, the advanced topics are discussed in Sections~\ref{lagrangian}
-- \ref{Lattice}.

\section{Monopoles as Classical Solutions}\label{monopoles}

In this section, which is pure pedagogical in nature we summarize the facts
known about the magnetic monopoles viewed as solutions to the classical
equations of motion. We consider both electrodynamics and gluodynamics.

\subsection{Dirac Veto, Dirac Quantization Condition}

As is mentioned in the Introduction, there is a number of constraints
imposed on the theory to ensure that the Dirac string does not produce any
physical effect.  First, there is the {\it Dirac veto} which forbids any
direct interaction with the string.  The condition is not trivial at all, in
fact. Indeed, let us consider perturbation theory and the scattering of a
charged particle off the string. In perturbative approach one uses the basis
of plane waves. Which implies that the particles can be found at any
space-point and their wave functions overlap with the string. Which is in
violation of the Dirac veto. Thus, a simple minded Born approximation can be
misleading. Of course, this consideration alone does not yet allow to judge
whether this overlap is significant numerically.  But a little bit more
involved estimates do indicate that it is important and the perturbation
theory can and does bring wrong results. We shall examine these issues in
much more detail in Sections~\ref{monoprenorm}, \ref{nabmonrenorm} devoted
to the radiative corrections and here just state that a consistent treatment
of the Dirac string is always non-perturbative in nature because of the
Dirac veto. This simple observation allows to resolve many theoretical
puzzles.

The best known constraint on the Dirac string seems to be {\it the Dirac
quantization condition} which ensures absence of the Aharonov-Bohm effect
for the electrons scattered off the string:

\beq
Q_e \oint_{string} {\bf A} \mathrm{d} {\bf x}~=~ Q_e\,Q_m ~=~ 2\pi k\,,
\label{quant}
\eeq
where $Q_e$ is the electric charge of the electron and $k$ is an integer
number.  Note that as a result of the quantization condition the interaction
of the monopoles is described by a large coupling constant if the electrons
interact weakly and vice versa.  Also, the interaction of an electron with
the monopole is always governed by a coupling of order unit and is never
weak. This is another reason for the use of non-perturbative methods in the
theory of monopoles, which is of course always a challenge.

\subsection{Energy of the Dirac String}

Unlike the quantization condition (\ref{quant}), the energy of the Dirac
string is rarely discussed in the literature.  It is worth to emphasize,
therefore, that naively the energy of the string is infinite in the
ultraviolet.

Indeed, let us estimate the energy of the string:
\beq
\epsilon_{string}\sim \int (\bh^{Str})^2 \, \mathrm{d}^3r \,\sim\,
{(Length)\over Q_e^2(Area)} \,\sim\,Q_e^{-2} \Lambda_{UV}^2 (Length)~~.
\label{quadratic}
\eeq
We accounted here for the fact that the magnetic flux is quantized (see
above), $$(flux)~\equiv~\int{\bf H}\cdot d{\bf s}~=~{2\pi k\over Q_e}$$ and
tended the cross section of the string denoted by $(Area)$ to zero at the
end of the calculation.  Thus, we substituted $(Area)^{-1}$ by
$\Lambda_{UV}^2$.

The radial part of the magnetic field is also associated
with an infinite energy:
\beq
\epsilon_{rad}~\sim\int (\bh^{Coul})^2 \,\mathrm{d}^3 r~\sim~{1\over
r_0}~\sim~\Lambda_{UV}\,.  \label{linear} \eeq
Note that this ultraviolet divergence is linear, i.e. weaker than the
divergence due to the string, see Eq.~(\ref{quadratic}).

The infinite magnetic field of the string may have more subtle
manifestations as well. Consider interaction of two magnetic monopoles with
magnetic charge $\pm Q_m$ placed at distance $R$ from each other. Then, by
the analogy with the case of two electric charges, we would like to have
the following expression for the interaction energy:
\beq
\label{right}
\epsilon_{int}~=~\int \bh_{1}^{Coul}\bh_{2}^{Coul}  \,\mathrm{d}^3 r~=~
- \frac{Q_m^2}{4\pi}\,\frac{1}{R} \,.
\eeq
Note, however, that if we substitute the sum of the radial and string fields
for $\bh_{1,2}$, then we have an extra term in the interaction energy:
\beq
\label{wrong}
\tilde{\epsilon}_{int}~=~
\int \left(\bh_{1}^{Coul}\bh_{2}^{Str}+\bh_{1}^{Str}\bh_{2}^{Coul}
\right)\,\mathrm{d}^3 r  ~=~
+ 2\,\frac{Q_m^2}{4\pi}\,\frac{1}{R}\,.
\eeq
In other words, the account of the string field would flip the sign of the
interaction energy!  This contribution, although looks absolutely finite, is
of course a manifestation of the singular nature of the string magnetic
field, $|\bh^{Str}|\sim (Flux)/(Area)$. Note that the integral in
(\ref{wrong}) does not depend on the shape of the string.

To maintain the unphysical nature of the Dirac string we should use a
regularization scheme which would allow to get rid of these singularities.

\subsection{Lattice Regularization in Abelian Theories}
\label{latticereg}

Since the monopoles are naively having divergent energy (or action) in the
ultraviolet, the regularization is a crucial issue.  Moreover, we would like
to follow the lattice formulation since the monopoles are observed on the
lattice.  And for a good reason, as we shall immediately see.

Consider first the $U(1)$ case. As is emphasized in Ref.
\cite{Polyakov-compact-U1}, the lattice formulation implies that a Dirac
string which produces no Aharonov-Bohm scattering costs no action as well.
The reason is very simple. The lattice action is written originally in terms
of the contour integrals like (\ref{quant}) rather than field strength
$F_{\mu\nu}$:
\beq
S~=~ \sum\limits_{p} \,\mathrm{Re} \;\exp\{iQ_e \oint_{\diff p} A_{\mu}dx^{\mu} \}\,,
\eeq
where the sum is taken over all the plaquettes $p$. Thus, the condition (\ref{quant})
means absence, in the lattice formulation, of both the Aharonov-Bohm effect and the quadratic divergence
(\ref{quadratic}).
Moreover, it is straightforward to see that the interference term (\ref{wrong}) also vanishes.

Later, we will also discuss Dirac strings which
correspond to negative plaquettes in the lattice formulation.
Such strings are associated with the fundamental monopoles with
$Q_m=1$ in units of the Dirac quantization condition discussed above. The energy
of such Dirac strings is infinite
in the continuum limit, in agreement with the naive estimate (\ref{quadratic}).
The interference term (\ref{wrong}), however, disappears in the lattice formulation
in this case as well. Moreover, this observation is crucial to understand
the heavy monopole potential in the continuum limit.

The radial field, $\bh^{Coul}$ may also cause problems with infinite energy,
see (\ref{linear}). The lattice regularization is not much specific in that case,
however. The role of $r_0$ is simply played by the lattice spacing $a$.
Thus, the probability to find a monopole on the lattice is suppressed by
the action as:
\beq
\label{suppression}
e^{-S}~\sim~\exp (-const\cdot Q_e^{-2} \,L/a)\,,
\eeq
where $L$ is the length of the monopole trajectory, and the $Q_e^{-2}$
factor appears because of the Dirac quantization
condition (\ref{quant}) which relates the magnetic charge $Q_m$ to the inverse
electric charge.

Although the Eq.~(\ref{suppression}), at first sight, rules out monopoles as
physically significant excitations, the fate of the monopoles in the
$U(1)$ case depends in fact on the value of the charge $Q_e$. The point is that
the entropy factor, or the number of various
trajectories with the same length $L$ grows also exponentially with the length of the monopole
trajectory:
\beq
(Entropy)~\sim~\exp(+const' \cdot L/a)\,,
\eeq
where the $const'$ is a pure geometric factor. As a result for $Q_e\sim 1$ there is a
phase transition corresponding to the condensation of the monopoles.
This phase transition, which is well studied on the lattice, is the first
and striking example of importance of the UV regularization
in the non-perturbative sector.

\subsection{Classification of Monopoles in non-Abelian Theories}

From now on, we will discuss monopoles in unbroken non-Abelian gauge theories,
having in mind primarily gluodynamics, i.e. quantum chromodynamics
without dynamical quarks. Moreover, for the sake of simplicity we will
consider only the $SU(2)$ gauge group.

A natural starting point to consider monopoles in non-Abelian theories
is their classification. There are actually a few approaches to
the monopole classification and it is important to realize both similarities
and differences between them.

{\it The dynamical, or $U(1)$ classification.}
Within this approach~\cite{U1-classification}, one looks for monopole-like
solutions of the classical Yang-Mills equations.
Where by the ``monopole-like'' solutions one understands
potentials which fall off as $1/r$ at large $r$, see Eq. (\ref{dirac}).
The basic finding is that there are no specific non-Abelian
solutions and all the monopoles can be viewed as Abelian-like embedded
into the $SU(2)$ group. Moreover, using the gauge invariance one can
always choose the corresponding $U(1)$ group as, say,
 the rotation group around the
third direction in the color space. According to this classification,
the monopoles are characterized by their charge with respect to a $U(1)$
which are integer numbers:
\beq
|Q_m|~=~0, ~1,~2,~...~~~~.
\eeq

{\it The topological, or $Z_2$ classification.}
The $Z_2$ classification~\cite{Z2-classification} is based entirely on topological arguments.
Namely, independent types of monopoles
are enumerated by considering the first homotopy group of the gauge group.
The $SU(2)$ gauge group is trivial since $\pi_1(SU(2))=0$, while in case of
$SO(3)$,
\beq
\pi_1(SU(2)/Z_2)~=~Z_2\label{z2},
\eeq
and there exists a single non-trivial topological monopole. We will denote the magnetic charge
of such monopoles as $|Q_m|=1$. Note, however, that the charges
$Q_m=\pm 1$ are indistinguishable in fact. As for the charges $Q_m=2$
they are equivalent, from this point of view, to no magnetic charge
at all.

The topological classification (\ref{z2}) is readily understood
as a classification of the Dirac strings whose end points
represent the monopoles.
Then there is only one non-trivial string, that is the one for which
Eq. (\ref{quant}) is satisfied for gluons but not for quarks.
Namely, because the $U(1)$ charge associated with gluons is twice as large as
that of the quarks
we may have
\beq
\exp\{\, ig \oint A_{\mu}dx^{\mu} \,\} ~=~ -1
\eeq
and such a string is still not visible for the isospin-one particles.
On the other hand, the standard plaquette action is based on the phase factor evaluated
for particles in the fundamental representation. Which means, in turn, that
the Dirac string is piercing the negative plaquettes.

\subsection{$Z_2$ Monopoles}\label{zeroactmon}

In principle, the $U(1)$ and $Z_2$ classifications are different.  Indeed,
while the $U(1)$ classification allows for any integer charge, the $Z_2$
classification leaves space only for a single non-trivial charge:
\beq
Q_m~=~0,1.
\eeq
The reconciliation of the two classifications is that the $U(1)$ solutions
with $|Q_m|\ge 2$ are in fact unstable because of the presence of massless
charged vector particles (gluons)~\cite{bn}. The instability of the
solutions implies that even if an external source with $|Q_m|\ge 2$ were
introduced into the vacuum state of the gluodynamics, charged gluons would
fall onto the monopole center because of the strong magnetic interactions.
Moreover, one can imagine that as result of this instability the charged
fields $A^{\pm}$ are build up as well.

In a somewhat related way, one can demonstrate
the apparent irrelevance of the $|Q_m|=2$ monopoles
by producing an explicit non-Abelian field configuration
which looks as a $|Q_m|=2$ monopole in its Abelian part
but has {\it no $SU(2)$ action} at all~\cite{main}. This field
configuration is a Dirac string with open ends,
which correspond to a monopole-anti-monopole pair separated by
distance $R$.
Note that the Abelian flux is still transported along the Dirac string
and is still conserved for the radial field.
What is lost, however, is the relation between the Abelian flux and action.
In the Abelian case non-vanishing flux means non-vanishing
magnetic field and non-vanishing action since the action density
is simply $\bh^2$. Now the action is $(F_{\mu\nu}^a)^2$ and the Abelian part
of the $F_{\mu\nu}^3$ can be canceled by the commutator term.

The simplest example illustrating the cancellation between the Abelian component
and the commutator term in $F_{\mu\nu}^a$ is produced by the potential:
\beq  \label{zeroA}
A_{\mu}^a~=~-{2\over r^2}\epsilon^{abc}r^b{\sigma^a\over 2}~~,\label{empty}
\eeq
where $\sigma^a$ are the Pauli matrices and $r^a$ is the radius vector.
One can readily check by a direct calculation that the corresponding
non-Abelian field-strength tensor vanishes identically (some care should be
exercised to analyze the singularity at $r=0$).  Another way to convince
oneself that the configuration (\ref{empty}) has zero action is to observe
that
\beq \label{OcrosO}
A_{\mu}^a~=~i(\Omega^0)^{\dagger}\partial_{\mu}\Omega^0,
\eeq
where
$$\Omega^0~=~i\sigma^an^a,$$
and $n^a$ is the unit vector looking from the center of the ``monopole''
to the observation point.

We pause here to note that the potential (\ref{empty}) does not exhibit any
Dirac string.  The reason is that we have not gauge rotated it to the form
suitable for the $U(1)$ classification of the monopoles. The absence of the
Dirac string may look appealing at first sight.  On the other hand, a
shortcoming of using an explicitly non-Abelian field, like (\ref{empty}) is
that one cannot add up in this case the monopole fields if the monopoles are
situated at different space points.  Thus, we are not using such forms in
this review although one cannot rule out that in future some progress can be
made to overcome the difficulty with adding up the monopoles written in the
form similar to (\ref{empty}).

Coming back to the problem of writing a
filed configuration which in its Abelian part
looks as a monopole-antimonopole pair ($|Q_m|=2$)
we will use formulation with a Dirac string.
Algebraically the problem becomes more complicated and here we reproduce
only the final answer~\cite{main}.
Namely, such a configuration is generated
from the vacuum
by the following gauge rotation matrix:
\beq
\label{example}
\Omega~=~\left(\matrix{
e^{i\varphi}\sqrt{A_D} & \sqrt{1-A_D}\cr
-\sqrt{1-A_D}          &  e^{-i\varphi}\sqrt{A_D}\cr
}\right)\,,
\eeq
where $\varphi$ is the angle of rotation around the axis connecting
the monopoles  and $A_D$ is the $U(1)$ potential representing pure Abelian monopole pair:
\beq
A_{\mu}dx_{\mu}~=~
{1\over 2}\left({z_+\over r_+}-{z_-\over r_-}\right)d\varphi
~\equiv~ A_D(z,\rho)d\varphi\,,
\eeq
where $z_{\pm}=z\pm R/2$, $\rho^2=x^2+y^2$, $r_{\pm}^2=z_{\pm}^2+\rho^2$.
Note that the action associated with the Dirac string is considered in this
case zero, in accordance with  the lattice version of the theory
(for details see~\cite{main}).

In this example, the monopoles with $|Q_m|=2$ are a kind of a pure
gauge field configurations carrying no action.

It is somewhat more difficult to visualize dynamically the equivalence of
the $Q_m=\pm 1$ monopoles, also implied by the $Z_2$ classification.
The mechanism mixing the $Q_m=\pm 1$ solutions seems to be the following.
Imagine that we start with, say, $Q_m=+1$ solution. Then a Dirac
string carrying the flux corresponding to the $Q_m=-2$ can be superimposed
on this solution. It is important at this point that
such a Dirac string costs no action (or energy). Then the radial
magnetic field can also change its direction since this does not
contradict the flux conservation any longer. In a related language,
one could say that the $|Q_m|=2$ monopoles are condensed in the
vacuum and as a result the magnetic charge can be changed freely
by two units.

As far as interaction of two $|Q_m|=1$ monopoles is concerned, one
might expect that they would interact as a monopole-antimonopole pair
rather than as a monopole-monopole configuration.  Indeed, the
monopole and antimonopole attract each other and thus represent the
lowest energy state of the system.

\subsection{Conclusions  \# 1}

Thus, the physics of the monopoles in non-Abelian theories in the classical approximation
turns very simple.

Namely, there exist only monopoles with $|Q_m|=1\equiv 2\pi/g$
where $g$ is the coupling constant of the non-Abelian $SU(2)$
theory.
The monopoles are infinitely heavy
and their interaction is
Abelian like:
\beq
V_{m\bar{m}}~=~-{Q_m^2\over 4\pi R}~=~-{\pi\over g^2 R}\,,
\label{potent}
\eeq
where $R$ is the separation between the monopoles.

Clearly enough, this first approximation falls far beyond
an adequate description of the empirical data on the monopoles,
see the Introduction. Thus, we are invited to go into more advanced approaches
which we would try to introduce step by step.

\section{Lattice Monopoles, Introductory Remarks}
\label{latticemonopoles}

\subsection{Dirac Strings on the Lattice, Non-Abelian Case}

The Dirac strings remain essentially Abelian objects in case of non-Abelian theories as well.
A novel feature is that the field transported along the string can be arbitrarily oriented in the
color space. Which is a kind of a new gauge freedom.

In more detail, the general one-plaquette action of $SU(2)$ lattice gauge
theory (LGT) can be represented as:
\beq
S_{lat}(U)~=~\beta
\,\sum\limits_p \, S_p\left(1-{1\over 2}TrU[\partial
p]\right) , \label{plaquette}
\eeq
where $\beta = 4/g^2$, $g$ is the bare coupling, $\partial
p$ is the boundary of an elementary plaquette $p$, the sum is taken over all
$p$, $U[\partial p]$ is the ordered product of link variables $U_l$ along
$\partial p$.  In particular, if $S_P(x)=x$ then (\ref{plaquette}) is the
standard Wilson action.  The exponent of the lattice field strength tensor
$F_p$ is defined in terms of $U[\partial p]$:
\beq
U[\partial p]~=~e^{i\hat{F}_p}~=~ \cos\big[{1\over
2}|F_p|\big]+i\tau^an^a\sin\big[{1\over 2}|F_p|\big],\label{partial}
\eeq
where $\hat{F}=F^a\cdot \tau^a/2, |F|=\sqrt{F^aF^a}$ and we define
$n^a_p~=~F_p^a/|F_p|$ for $|F_p|\neq 0$, while $n^a_p$ is an arbitrary unit
vector for $|F_p|=0$.

The lattice action (\ref{plaquette})
depends only on $\cos\big[{1\over 2}|F_p|\big]$. Therefore the action
of the $SU(2)$ LGT possesses not only the usual gauge symmetry, but allows
also for the gauge transformations which shift the field strength tensor
by $4\pi k$, $|F_p|\to |F_p|+4\pi k$,
$k\in Z$:
\beq
e^{i\hat{F}_p} = \exp\{i|F_p|\hat{n}_p\}
 = \exp\{i(|F_p|+4\pi)\hat{n}_p\} =
\exp\{i(F_p^a+4\pi n^a_p)\tau^a/2\}.
\eeq
Thus, the symmetry inherent to the lattice formulation can be represented as:
\beq
F_p^a~\to~F_p^a+4\pi n_p^a,~~~\vec{F}_p\times \vec{n}_p~=~0,~~~n_p^2=1.
\label{symmetry}\eeq
The symmetry (\ref{symmetry}) is absent in the conventional continuum
limit, $\int(F_{\mu\nu}^a)^2d^4x$. If $n_p\neq 0$ then one can say that
a Dirac string is piercing the plaquette. Similar to the Abelian case,
such a Dirac string costs no action.

Note that in the continuum limit $n_p^a$ becomes a singular
two-dimensional structure $\dual{\Sigma}^a_{\mu\nu}$ which is
representing the Dirac string world sheet.

\subsection{The 't~Hooft Loop on the Lattice} \label{Hlooplat}

So far we discussed invisible Dirac strings. As we know, the Dirac
strings corresponding to the fundamental monopoles with $Q_m=1$ have
an infinite action, or energy. In the lattice formulation they
correspond to the plaquettes with the phase factor $exp(i\pi)$.  This
infinite energy precludes the fundamental monopoles from being
dynamical objects present in the QCD vacuum. However, they can be
introduced as external objects. The infinite energy of the
corresponding Dirac string then does not matter, the same as with the
infinite self-energy of heavy quarks introduced via the Wilson loop.
The fundamental monopoles can be introduced on the lattice via the
't~Hooft loop~\cite{loop}.  In a simplified way, the monopoles are
visualized as end-points of the corresponding Dirac strings which in
turn are defined as piercing negative plaquettes.  The trick to
introduce the negative plaquettes on the lattice is to formally change
the sign of the coupling $g^2$ on a manifold of plaquettes. Then these
plaquettes become negative in the limit $|g^2|\to 0$.

Proceeding to more detailed definitions, the 't~Hooft loop is
formulated~\cite{susskind} in terms of the action
\beq
S(\beta,- \beta) =
\beta \sum\limits_{p\notin M} \;{\mathrm Tr}\;U_p
- \beta\sum\limits_{p\in M}\;{\mathrm Tr}\;U_p\,,
\label{S1}
\eeq
where $M$ is a manifold which is dual to the surface spanned on the monopole
world-line $j$.  Introducing the corresponding partition function,
$Z(\beta,-\beta)$ and considering a planar rectangular $T\times R$, $T\gg R$
contour $j$ one can define
\beq
V_{m\bar{m}}(R) ~\equiv ~ -\frac{1}{T} 
\ln{ Z(\beta,-\beta)\over Z(\beta,\beta)}\,.
\label{energy}
\eeq
By the analogy with expectation value of the Wilson loop
the quantity $V_{m\bar{m}}(R)$ is referred to as monopole-antimonopole,
or heavy monopole
potential.
The potential $V_{m\bar{m}}(R)$, in a way, is the same fundamental
quantity as the heavy-quark potential $V_{Q\bar{Q}}$
and its understanding within the fundamental QCD would be
of great importance.

\subsection{Lattice Monopoles in the Abelian Projection of
$SU(2)$ Gluodynamics} \label{DefLatMon}

The $SU(2)$ gauge fields $U_l$ on the lattice are defined by $SU(2)$
matrices  attached to the links $l$. These lattice fields are
related to the continuum $SU(2)$ fields $\hat A$: $U_{x,\mu} =
e^{i a g \hat{A}_\mu (x)}$, here $a$ is the lattice spacing. Under
the gauge transformation,  the field $U_l$ transforms as
$U_{x,\mu}^\Omega = \Omega_x^+ \, U_{x,\mu} \, \Omega_{x+\hat\mu}$,
the matrices of the gauge transformation are attached to the sites
$x$ of the lattice.

The Maximal Abelian gauge is defined on the lattice by the following
condition~\cite{MAAP}:
\beq \label{MAAPdef}
\max_\Omega R[\hat{U}^\Omega_l]\,,\qquad
R[U_l] = \sum\limits_l Tr[\sigma_3 U_l^+ \sigma_3 U_l]\,,
\quad l = \{x,\mu\}\,.
\eeq
This gauge condition corresponds to an Abelian gauge, since $R$ is
invariant under the gauge transformations defined by the matrices:
\beqn
\Omega_{U(1)} (x) = \pmatrix{
e^{i\alpha(x)} & 0 \cr 0 & e^{-i\alpha(x)} \cr}\,, \quad
\alpha\in[0,2\pi)\,.
\label{U1}
\eeqn

Let us parameterize the link matrix $U$ in the standard way \beqn
\label{Uparameters} U_l = \pmatrix{ \cos \varphi_l e^{i\theta_l} & \sin
\varphi_l e^{i \chi_l}\cr - \sin \varphi_l e^{- i \chi_l} & \cos \varphi_l
e^{-i\theta_l} \cr}\,, \label{LatticeU}
\eeqn
where $\theta,\chi\in[-\pi,+\pi)$ and $\varphi\in[0,\pi)$. In this
parameterization,
\beqn \label{Rphi}
R[U_l] = \sum_l \cos 2 \varphi_l\,.
\eeqn
Thus, the maximization of $R$ corresponds to the maximization of
the diagonal elements of the link matrix~\eq{LatticeU}.

Under the $U(1)$ gauge transformations, \eq{U1},
the components of the gauge field \eq{LatticeU} are transformed as
\beqn
\theta_{x,\mu} \rightarrow \theta_{x,\mu}
+  \alpha_x - \alpha_{x+\hat\mu},\,\,
\chi_{x,\mu} \rightarrow \chi_{x,\mu}
+  \alpha_x + \alpha_{x+\hat\mu}, \,\,
\varphi_{x,\mu} \rightarrow \varphi_{x,\mu}.
\eeqn
Therefore,  in the MaA the gauge,  the field $\theta$ is the $U(1)$ gauge
field, the field $\chi$ is the Abelian charge 2 vector matter
field, the field $\varphi$ is the non--charged vector matter field.

\begin{figure}[htb]
\vskip5mm
\begin{center}
\begin{tabular}{cc}
{\epsfxsize=0.45\textwidth\epsfbox{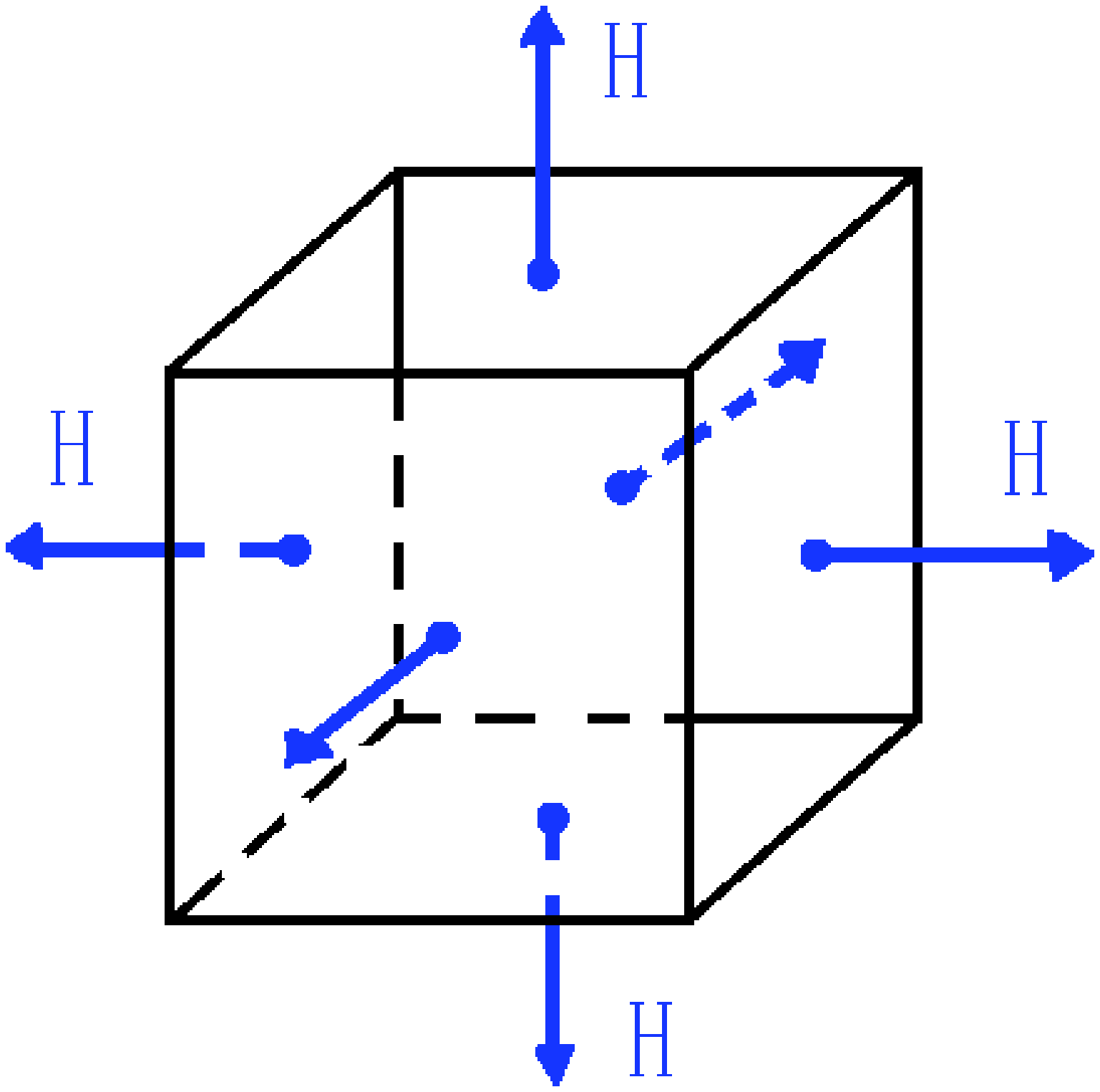}}& \vspace{.2cm}
{\epsfxsize=0.25\textwidth\epsfbox{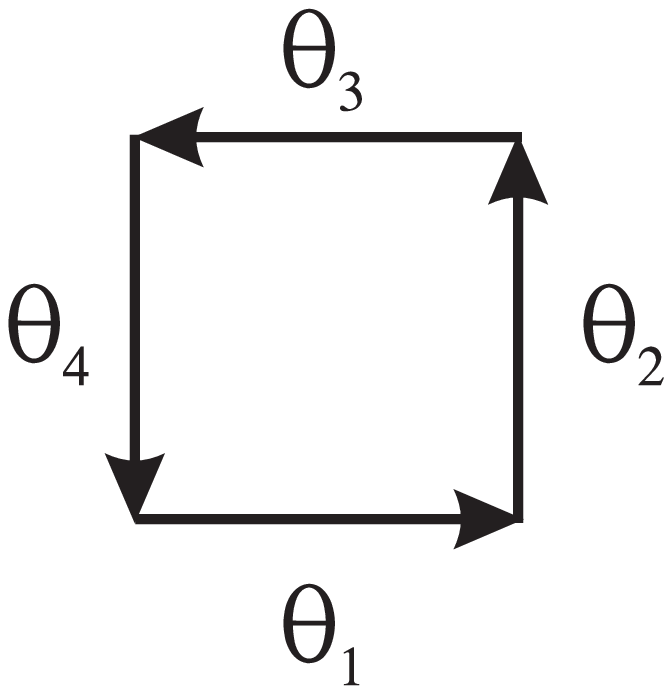}}\\
 (a) & (b)\\
\end{tabular}
\end{center}
\caption{The magnetic flux through the boundary of the
cube $C$ (a) and lattice field strength tensor (b).}
\label{cube0}
\end{figure}

A  configuration of Abelian gauge fields $\theta_l$ can
contain monopoles. The position of the monopoles is defined by the
lattice analogue of the Gauss theorem. Consider the elementary
three--dimensional cube $C$ (Figure~\ref{cube0}(a)) on the lattice.

The Abelian magnetic flux $\vec{H}$ through the surface of the cube
$C$ is given by the formula

\beqn
m = \frac{1}{2 \pi}
\sum\limits_{P\in\partial C} {\bar \theta}_P\,, \label{LatticeFlux}
\eeqn
where ${\bar \theta}_P$ is the magnetic field defined as
follows.  Consider the plaquette angle $\theta_P = \theta_1 + \theta_2
- \theta_3 - \theta_4 \equiv \dd \theta$, the $\theta_i$'s are
attached to the links $i$ which form the boundary of the plaquette
$P$, Figure~\ref{cube0}(b). The definition of ${\bar \theta}_P$ is
${\bar \theta}_P = \theta_P + 2 \pi k$, where the integer $k$ is such
that $- \pi < {\bar \theta}_P \le \pi$. The restriction of ${\bar
\theta}_P$ to the interval $(-\pi,\pi]$ is natural since the Abelian
action for the compact fields $\theta_l$ is a periodic function of
${\bar \theta}_P$. Equation~\eq{LatticeFlux} is the lattice analogue
of the continuum formula $m = \oint \vec{H} \, {\rm d} \vec{S}$. Due
to the compactness of the lattice field $\theta$ ($-\pi < \theta \le
\pi$) there exist singularities (Dirac strings), and therefore,
$\mbox{div}\vec{H} \neq 0$.  \vskip3mm

The magnetic charge $m$ defined by eq.~\eq{LatticeFlux} has the
following properties:
\begin{enumerate}
\item $m$ is quantized: $m = 0,\pm 1,\pm 2$;
\item If $m \neq 0$,  then there exists a magnetic current $j$. This
current is attached to the link  dual to cube $C$.
\item Monopole currents $j$ are conserved:  $\delta j = 0$, the
currents form closed loops on the $4D$ lattice.
\end{enumerate}

We discuss in details the properties of the lattice monopoles in
gluodynamics in Section~\ref{Lattice}.

\subsection{Dirac Strings and Singular Gauge Transformations}

If we wish to build up the continuum theory in a way reproducing the basic
features of the lattice formulation, we need first of all to ensure the
invisibility of the Dirac string.  One way is to put the energy of the Dirac
strings to zero ``by hands''.  A more educated way is to allow for singular
gauge transformations which would correspond to introduction of the Dirac
strings.  Let us start with the compact electrodynamics (cQED).

Under the gauge transformation, $A_\mu \to
A_\mu + \partial_\mu \alpha$, the field strength tensor transforms
as: $F_{\mu\nu} \to F_{\mu\nu} + [\partial_\mu,\partial_\nu]\alpha$.
For a singular gauge transformation
$[\partial_\mu,\partial_\nu]\alpha\neq 0$. Moreover,
for a sufficiently general singular
gauge parameters $\alpha$ the commutator of derivatives is
proportional to the $2D$ $\delta$ function on some closed surface:
$[\partial_\mu,\partial_\nu]\alpha \propto
{}^*\Sigma_{\mu\nu}(\alpha)$. If $x$ are the $4D$ coordinates of the
surface, $\sigma_\alpha$ ($\alpha =1,2$) are $2D$ coordinates on the
world-sheet, then the general representation of the surface is:
\beqn
\Sigma_{\mu\nu} = \int d^2\sigma \sqrt{g} \; t_{\mu\nu}(\sigma) \;
\delta^{(4)}(x-\tilde{x}(\sigma))\,, \,\,
g(\sigma)~=~\mathrm{Det}[\; \diff_\alpha \tilde{x}_\mu \;
\diff_\beta \tilde{x}_\mu \;]\, , \\
t_{\mu\nu}(\sigma) =  {1\over \sqrt{g}} \;
\varepsilon^{\alpha\beta} \;
\diff_\alpha \tilde{x}_\mu \; \diff_\beta \tilde{x}_\nu\,,\,\,
t^2_{\mu\nu} = 2\, .
\eeqn
Now, we can write down a partition function of electrodynamics which is
invariant under the singular gauge transformations:
\beq\label{ZcQED}
\cZ~=~\int \cD A_\mu \cD \Sigma \exp\left\{ - \frac{1}{2e^2}
\int \left[ F_{\mu\nu} +
2\pi {}^*\Sigma_{\mu\nu}\right]^2 \right\}\, .
\eeq
Of course $\int \, \cD \Sigma$ is only formal notation for the
summation over all surfaces in $4D$ as far as we do not specify the
measure. On the other hand, this summation is well defined on the
lattice and \eq{ZcQED} can be viewed as the partition function of cQED
with the Villain type of the action written in the continuous
notations, if $\cD \Sigma$ in \eq{ZcQED} includes summation over closed
and not closed surfaces.

Proceeding now to the non-Abelian case, we note again that the
symmetry (\ref{symmetry}) is absent in the conventional continuum
action, $\int (F^a_{\mu\nu})^2\, d^4 x$.  To bring the continuum
theory into agreement with the lattice formulation, we need a
generalization of (\ref{ZcQED}).  In the continuum limit $n^a_p$
becomes a singular two-dimensional structure $\dual \Sigma^a_{\mu\nu}
= \frac{1}{2} \varepsilon_{\mu\nu\lambda\rho} \Sigma^a_{\lambda\rho}$
which is a generalization of the Dirac strings in the compact
electrodynamics and which transforms in the adjoint representation of
the gauge group:
\beq
\label{Sigma-colored-1zz}
\Sigma^a_{\mu\nu} = \int d^2\sigma \sqrt{g} \; t^a_{\mu\nu}(\sigma) \;
\delta^{(4)}(x-\tilde{x}(\sigma))\,.
\eeq
The surface (\ref{Sigma-colored-1zz})
need not to be closed.
Moreover, the second equality in (\ref{symmetry}) requires that
\beq
\label{constraint-1}
\vec{t}_{\mu\nu}(\sigma) \times \dual{\vec{F}}_{\mu\nu}(\tilde{x}) = 0\,,
\eeq
where the continuum field strength tensor
${\hat F}_{\mu\nu} = \diff_\mu {\hat A}_\nu -
\diff_\nu {\hat A}_\mu - i [{\hat A}_\mu, {\hat A}_\nu]$.
Eq.~(\ref{constraint-1})
determines the color structure of $t^a_{\mu\nu}$:
\beq
\label{tn}
t^a_{\mu\nu}(\sigma) ~=~ t_{\mu\nu}(\sigma) ~ n^a(\sigma) \, ,
\qquad
n^a(\sigma) ~=~ (t \cdot \dual{F}^a ) \left[{(t \cdot \dual{F}^b )^2}\right]^{-1/2}
\,,
\eeq
where $(t \cdot F^a) \equiv t_{\mu\nu}(\sigma) \;
F^a_{\mu\nu}(\tilde{x})$ and $n^a$ is normalized as $\vec{n}^2=1$. On
the set of points where $(t \cdot \dual{F}^a) = 0$ the direction of
$n^a(\sigma)$ is arbitrary.

Now, the continuum analog of the lattice symmetry (\ref{symmetry}) is:
\beq
\label{discrete-symmetry}
F^a_{\mu\nu} ~ \to ~ F^a_{\mu\nu} + 4\pi \; \dual{\Sigma}^a_{\mu\nu}\,,
\eeq
The action of $SU(2)$ gluodynamics which possesses the additional symmetry
(\ref{discrete-symmetry}) can be formally represented as:  \beq
\label{part-fun-general}
Z= \int \cD A\;\exp\Bigl\{ - S(F) \Bigr\}\,,
\eeq
\beq
S(F) = - \log \; \int \cD\Sigma \exp\Bigl\{ - {1\over 4 g^2} \int d^4 x \;
\Bigl[ F^a_{\mu\nu} + 4\pi\; \dual{\Sigma}^a_{\mu\nu} \Bigr]^2
\Bigr\}\,,
\label{new-action}
\eeq
where the integration is over all possible surfaces,
\beq
\label{Sigmacolor}
\Sigma^a_{\mu\nu} = \int d^2\sigma_{\mu\nu} \; n^a(\sigma) \;
\delta^{(4)}(x-\tilde{x}(\sigma))\, .
\eeq

The expressions (\ref{part-fun-general},\ref{new-action}) are only formal
since it is impossible to separate rigorously the measure $\cD\Sigma$ from
the gauge degrees of freedom in $\cD A$.  Nevertheless, the
Eq.~(\ref{part-fun-general},\ref{new-action}) is a good starting point since
it reproduces the basic symmetries of the theory.  Indeed, the action
(\ref{new-action}) is invariant under smooth $SU(2)$ gauge transformations
since vector $n^a$ transforms in the same way as $F^a_{\mu\nu}$ does.  It is
also invariant under transformations (\ref{discrete-symmetry}) which
correspond to the lattice symmetry relations (\ref{symmetry}).

Note also that for self-intersecting surface  $\Sigma_{\mu\nu}$,
Eq.~(\ref{Sigma-colored}), the world-sheet vector field $n^a(\sigma)$ is
generally multi-valued as function of $\tilde{x}$. Furthermore, for the
non-orientable surfaces the field $n^a(\sigma)$ cannot be defined smoothly
everywhere on $\Sigma$. To avoid these complications we consider only the
orientable surfaces without self-intersections. This reservation is specific
for Dirac strings in the non-Abelian case.

\subsection{The 't~Hooft Loop in the Continuum Limit}

After the experience with working out the continuum limit of the Dirac string world sheet
in the preceding subsection, there is no difficulty to figure out that
a continuum analog of the 't~Hooft loop looks as~\cite{main}:
\beq
\label{tHooft-general}
H(\Sigma_\cC) =
\exp\left\{
{1\over 4g^2}\int d^4 x \; \left[
\left(F^a_{\mu\nu}\right)^2
-
\left(F^a_{\mu\nu} + 2 \pi \dual{\Sigma}^a_{\cC \; \mu\nu}\right)^2
\right]\right\}\,,
\eeq
\beq
\label{Sigma-colored}
\Sigma^a_{\cC\;\mu\nu} = \int d^2\sigma_{\mu\nu} \; n^a(\sigma) \;
\delta^{(4)}(x-\tilde{x}(\sigma))\,,
\eeq
where the surface $\Sigma^a_\cC$ spanned on the contour $\cC$ is
assumed to be non-intersecting.  The unit three-dimensional vector
field $n^a(\sigma)$, $\vec{n}^2=1$ is defined on the world-sheet of
the Dirac string. For analogous definitions see (\ref{tn}).  Therefore
$n^a(\sigma)$ is not an independent variable, it is completely
determined by the components of the field strength tensor
$F^a_{\mu\nu}$.  It can be shown~\cite{main} that the
Eq.~(\ref{tHooft-general})-(\ref{tn}) define the correct 't~Hooft loop
operator the expectation value of which depends only on the contour
$\cC$, not on a particular position of the surface $\Sigma_\cC$.

\subsection{Conclusions \#2}

The main lesson from the lattice formulation which we learnt so far is that
to match the lattice and continuum formulations of the gluodynamics one should allow
for singular gauge transformations. In this way one can hope to reproduce the
compactness of the $U(1)$ subgroups of the $SU(2)$ group and the
non-observability of the Dirac strings which is crucial for introduction of
the monopoles.

\section{Lagrangian Approach to the Dual Gluodynamics}
\label{lagrangian}

In this section we are going to derive some important results concerning
monopoles and their interaction in gluodynamics, understood as a continuum
limit of the lattice formulation. At first we discuss the Zwanziger receipt
of introduction of the dual photons (Subsection~\ref{dualphoton}). In
Subsection~\ref{dualgluon} we introduce a Zwanziger--type Lagrangian for
gluodynamics. The central point here is the match of ordinary gluons in an
adjoint representation with a single dual gluon which is an $U(1)$ Abelian
gauge boson. In Subsections~\ref{monoprenorm}, \ref{nabmonrenorm} we use the
Lagrangian formulation to derive the short distance behavior of the heavy
monopole potential, with one-loop quantum corrections included. In
Subsections~\ref{PhenLagr}, \ref{Casimir}  we discuss the phenomenological
Lagrangians describing the monopole condensation. The central point here is
a derivation of the Casimir scaling.

\subsection{Photons, Direct and Dual}
\label{dualphoton}

Now, that we know that magnetic monopoles are relevant degrees of freedom
on the lattice, at least in some gauges, we would like to have a continuum
version of gluodynamics which would allow both for color and magnetic charges.
We begin, however, with a review of the corresponding construction
in case of the electrodynamics, that is of the Zwanziger Lagrangian
\cite{zwanziger}.

We have already mentioned some basic features of the Zwanziger approach
in the Introduction and here would like to directly proceed to
the Lagrangian:
\beq
\label{Zw-action}
L_{Zw}(A,B)~=~ \frac{1}{2}(m\cdot[\diff\wedge A])^2 ~+~ \frac{1}{2}(m\cdot[\diff\wedge B])^2 ~+
\eeq
$$
+~\frac{i}{2}(m\cdot[\diff\wedge A])(m\cdot\dual{[\diff\wedge B]}) ~-~
\frac{i}{2}(m\cdot[\diff\wedge B])(m\cdot\dual{[\diff\wedge A]})
~+~ i\,j_e\cdot A ~+~ i\,j_m\cdot B\,,
$$
where $j_e,j_m$ are electric and magnetic
currents, respectively, $m_{\mu}$ is a constant vector, $m^2=1$ and
$$
[A\wedge B]_{\mu\nu} = A_\mu B_\nu - A_\nu B_\mu\,, \qquad
(m \cdot [A\wedge B])_\mu = m_\nu [A\wedge B]_{\mu\nu}\,,
$$
$$
\dual{[A\wedge B]}_{\mu\nu} ~=~ \frac{1}{2}\,\varepsilon_{\mu\nu\lambda\rho}\,
[A\wedge B]_{\lambda\rho}\,.
$$
At first sight, we have introduced two different vector fields, $A$ and $B$,
to describe interaction with electric and magnetic charges,
respectively. If it were so, however, we would have solved a wrong problem because
we need to have a single photon interacting both with electric
and magnetic charges. And this is what is achieved by the construct
(\ref{Zw-action}). Indeed, the action (\ref{Zw-action}) is not diagonal
in the $A$, $B$ fields and one can convince oneself that
the form of the $A,B$ interference terms in (\ref{Zw-action})
is such that the field strength tensors constructed on the potentials $A$ and $B$
are in fact related to each other:
\beq
m_{\mu}F_{\mu\nu}(A)~=~m_{\mu}\dual{F}_{\mu\nu}(B)~,\label{onetwo}
\eeq
where $\dual{F}$ denotes the dual tensor.
In perturbation theory, Eq. (\ref{onetwo}) ensures that there are only two physical degrees of freedom
corresponding to the transverse photons which can be
described either in terms of the potential $A$ or $B$.
Topological excitations, however, can be different in terms of $A$ and $B$.

The physical content of (\ref{Zw-action}) is revealed by the propagators
for the fields $A,B$.
In the $\alpha$-gauge one can derive:
\beqn
\label{propagators}
\langle A_{\mu} A_{\nu}\rangle ~=~
\langle B_{\mu} B_{\nu}\rangle ~=~
\frac{1}{k^2}\,(\delta_{\mu\nu}~-~(1-\alpha)\,\frac{k_\mu k_\nu}{k^2})\,,
\\
\\
\langle A_{\mu} B_{\nu}\rangle ~=~
- \langle B_{\mu} A_{\nu}\rangle ~=~
\frac{i}{k^2 (km)} \; \dual{[m\wedge k]}_{\mu\nu}\,.
\eeqn
The propagators should reproduce, as usual, the classical solutions. And indeed,
the $\langle AA \rangle$, $\langle BB \rangle$ propagators describe
the Coulomb-like interaction of two electric charges and magnetic monopoles, respectively.
While the $\langle AB \rangle$ propagator reproduces interaction of the magnetic field
of a monopole with a moving electric charge. The appearance of the poles in $(k\cdot m)$
is a manifestation of the Dirac strings.

To summarize, the Zwanziger Lagrangian in
electrodynamics~\cite{zwanziger} reproduces the classical interaction
of monopoles and charges. Upon the quantization, it describes the
correct number of the degrees of freedom associated with the photon.

\subsection{Monopole Charge Renormalization in Abelian Theory}
\label{monoprenorm}

We will now demonstrate that the quantum field theory with the
Zwanziger Lagrangian is not selfconsistent, and the knowledge of
propagators \eq{propagators} is not enough to get the correct
answers. To demonstrate these difficulties we consider the simplest
theory containing light electrons, photons, and magnetically charged
particles which are heavy and thus we can neglect their contributions
to the vacuum polarization.

Consider the renormalization of the monopole charge due to the vacuum
polarization induced by the electron loop, see the review
\cite{blagoevic} and references therein. As is emphasized in
Ref.~\cite{main}, the crucial point is that only loops with insertion
of two external (i.e., monopole) fields can be considered despite of
the fact that there is no perturbative expansion in $g\cdot e$.
Indeed, considering more insertions makes the graphs infrared
sensitive, with no possibility for $\ln\Lambda_{UV}$ to emerge.

\begin{figure}%[!htb]
% \begin{minipage}{11.5cm}
\begin{center}
 \epsfig{file=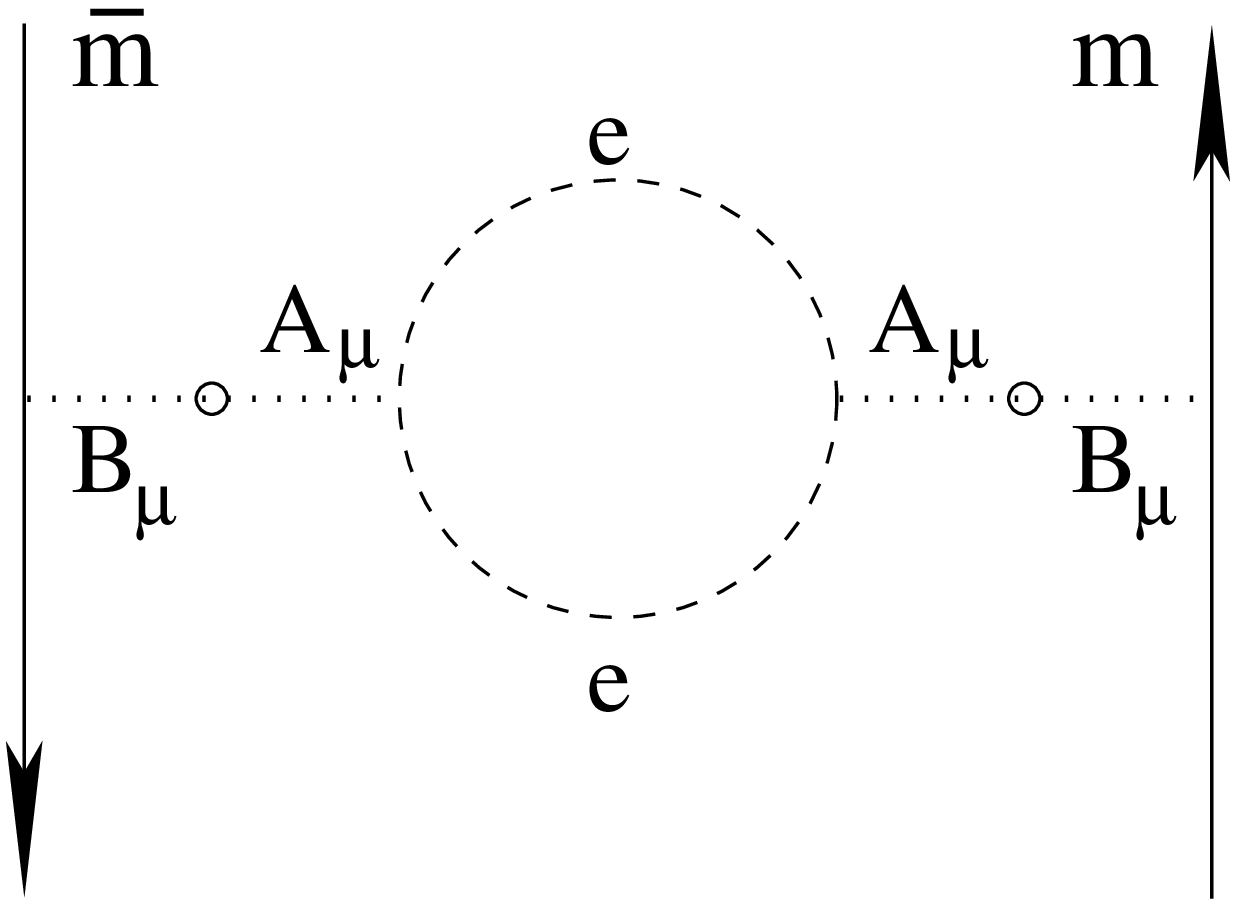,height=4.cm} \hspace{1.cm}
 \epsfig{file=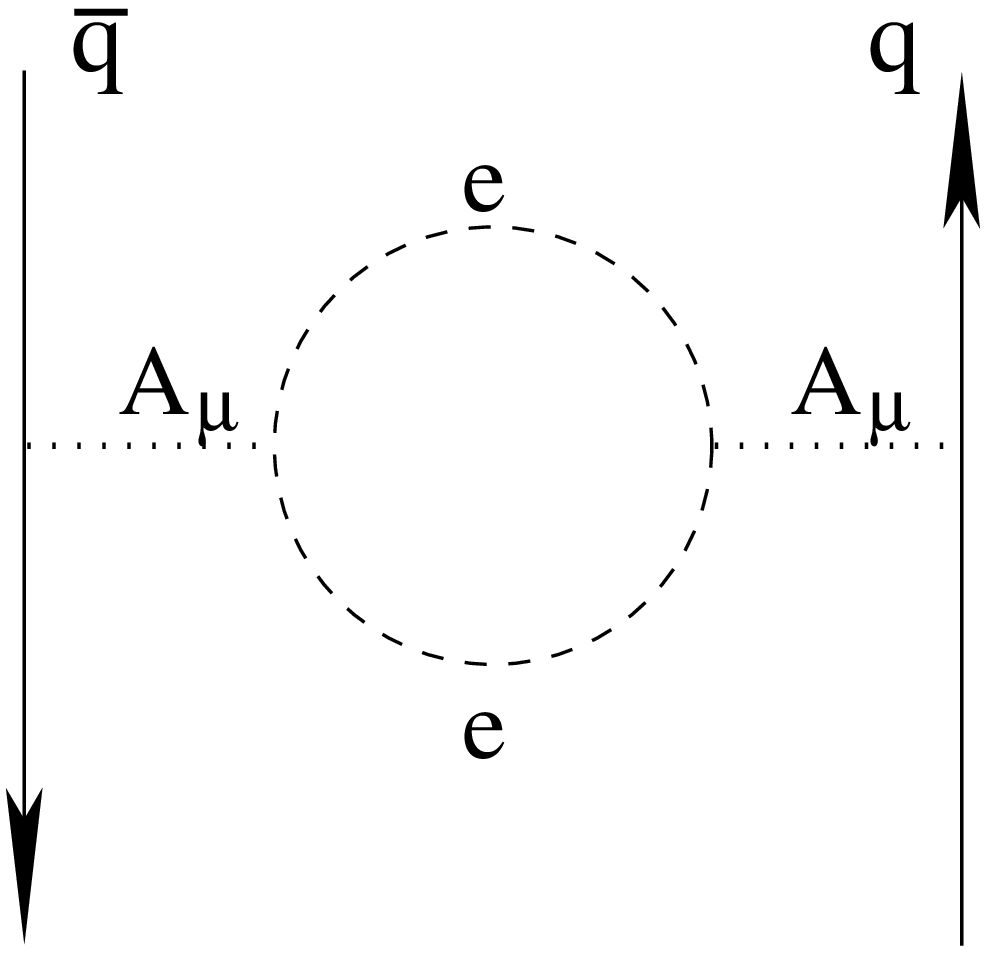,height=4.0cm}\\
\mbox{}\hspace{.6cm} (a) \hspace{5.5 cm} (b)\\
\end{center}
%\end{minipage}
\caption{
(a) Renormalization of the $A_\mu$ gauge field propagator.
(b) Renormalization of the $B_\mu$ gauge field propagator.}
\label{1}
\end{figure}

Then, the evaluation of, say, first radiative correction to the
propagator $\langle B_{\mu}B_{\nu}\rangle$ in the Zwanziger formalism
(\ref{propagators}) seems very straightforward and reduces to taking a
product of two $\langle AB \rangle$ propagators and inserting in
between the standard polarization operator of two electromagnetic
currents, see Fig. 1(a). The result is~\cite{blagoevic}:
\beq
\label{sch}
\langle B_{\mu}B_{\nu}\rangle(k) ~=~{\delta_{\mu\nu}\over k^2} (1-L)
~+~ {1\over (k\cdot m)^2}(\delta_{\mu\nu}-m_{\mu}m_{\nu}) L,
\eeq
where $$
L~=~{\alpha_{el}\over 6}\ln{\Lambda_{UV}^2/k^2}
$$
and we have neglected the electron mass so that the infrared cut-off
is provided, in the logarithmic approximation, by the momentum $k$.

At first sight, there is nothing disturbing about the result
(\ref{sch}). Indeed, we have a renormalization of the original
propagator which is to be absorbed into the running coupling, and a
new structure with the factor $(k\cdot m)^{-2}$ in front which is
non-vanishing, however, only on the Dirac string. The latter term
would correspond to the renormalization of the Dirac-string
self-energy which we do not follow in any case since it is included
into self-energy of the external monopoles. What is, actually,
disturbing is that according to (\ref{sch}) the magnetic coupling would
run exactly the same as the electric charge (see Fig. 1(b)),

\beq \label{AA}
\langle
A_{\mu}A_{\nu}\rangle(k) ~=~ (1-L){\delta_{\mu\nu}\over k^2}\,,
\eeq
violating the Dirac quantization condition.

The origin of the trouble is not difficult to figure out. Indeed,
using the propagator $\langle AB \rangle$ while evaluating the
radiative corrections is equivalent, of course, to using the full
potential corresponding to the Dirac monopole $A_D^{cl}$. Then,
switching on the interaction with electrons would bring terms like
$A_D^{cl}\, \bar{\psi} \gamma\psi$. Since $A_D^{cl}$ includes the
potential of the string, electrons do interact with the Dirac sheet and
we are violating the Dirac veto which forbids any direct interaction
with the string.

Let us demonstrate quantitatively that, indeed, it is the interaction
with the Dirac string that changes the sign of the radiative
correction. This can be done in fact in an amusingly simple
way. First, let us note that it is much simpler to remove the string
if one works in terms of the field strength tensor, not the
potential. Indeed, we have ${\bf H} = {\bf H}^{Coul} + {\bf
H}^{Str}$ while in terms of the potential $A_\mu$ any separation of
the string would be ambiguous.

We discussed already that the account of the string field would flip
the sign of the interaction energy, see eq. \eq{wrong}. On the
other hand the duality principle \eq{right} implies that at the
classical level we should write:

\beq
{\bf H}_1\cdot {\bf H}_2~\equiv~ {\bf H}_{1}^{Coul}\cdot {\bf
H}_{2}^{Coul}\,.
\eeq

However, the electrons in the diagram shown in Fig. 1(a) interact with
the full potential $A^{cl}_D$ and therefore the first radiative correction
would bring the product of the total ${\bf H}_1\cdot {\bf H}_2$ which
includes also the string contribution\footnote{ At this point we
assume in fact that $\Lambda_{UV}$ is larger than the inverse size of
the string, which is convenient for our purposes. Other limiting
procedures could be considered as well, however.}. Indeed, the result
in the log approximation would be as follows:

\beq
\label{delta}
\delta ({\bf H}_1\cdot {\bf H}_2)~=~L({\bf H}_{1}^{Str}+{\bf
H}_{1}^{Coul})\cdot ({\bf H}_{2}^{Str}+{\bf H}_{2}^{Coul})~=~-L {\bf
H}_{1}^{Coul}\cdot {\bf H}_{2}^{Coul}\,,
\eeq
where at the last step we have used the observation (\ref{wrong}).

Now, it is clear how we could ameliorate the situation. Namely, to keep the
Dirac string unphysical we should remove the string field from the
expression (\ref{delta}) which arises automatically if we use the
propagators (\ref{propagators}) following from the Zwanziger Lagrangian.
Thus, one should change ${\bf H}_1\cdot{\bf H}_2$ in the expression
(\ref{delta}) to ${\bf H}_{1}^{Coul} \cdot {\bf H}_{2}^{Coul}$.  The
 justification is that we should remove the effect of the string field from
any observable and consider this as a constraint on the ultraviolet
regularization. The constraint is satisfied by the lattice regularization.
The corrected propagator $\langle B_{\mu}B_{\nu}\rangle$ at the one-loop
level is becoming then:

\beq
\label{correct}
\langle B_{\mu}B_{\nu}\rangle(k) ~=~{\delta_{\mu\nu}\over k^2} (1+L)\,.
\eeq
In particular, there is no Dirac string self-energy renormalization.

The most important conclusion is that (\ref{correct}) does satisfy the Dirac
quantization condition also for the running coupling (compare eq. \eq{AA}
and eq. \eq{correct}).

It might worth emphasizing that the running of the magnetic coupling became
possible only because we do not have any longer the Maxwell equation
$div~{\bf H}=0$ on the operator level. Indeed, we have a source of the dual
photon (or gluon) which is mixed up with the ordinary photon (gluon).
Similarly, for the electric field we have $div~{\bf E}=\rho_{el}$.
Classically, for a point-like charge $\rho_{el}=Q_e\delta^{(3)}({\bf r})$ and
the coupling is not running of course.  However, on the quantum level the
fact that $div{\bf E}\neq 0$ as far as the operator equations are concerned
gets revealed through the running of the electrical coupling. Now the story
repeats itself, in the dual formulation, for the magnetic coupling. In
Subsection~\ref{nabmonrenorm} we demonstrate that in the non-Abelian case the
Dirac quantization condition is also valid for the running coupling.

\subsection{Dual Gluon as an Abelian Gauge Boson}
\label{dualgluon}

Now, if we try to extend the Lagrangian approach to
the dual gluons, we immediately come to a paradoxical conclusion that the
dual field, if any, is Abelian. Indeed, we have already
emphasized that monopoles associated with, say, $SU(N)$ gauge group
are classified according to $U(1)^{N-1}$ subgroups~\cite{U1-classification}
and can be realized as pure Abelian objects. Thus, there is no place for a
non-Abelian dual gluon because the monopoles do not constitute representations
of the non-Abelian group.

From this simple observation we immediately conclude that the dual field $B$
is an Abelian gauge field. Therefore in the Lagrangian  a la Zwanziger the
dual field $B$ can be mixed up only with one of the fields $A^a$.
Moreover, which field $A^a$ is being mixed up is  matter of a gauge fixing.
The Lagrangian realizing these ideas looks as 
(for more detail see~\cite{main}):
\beq
\label{dualglue}
{ L_{\mathrm{dual}}(A^a,B) ~=~ }
%&
	\frac{1}{4}( F^a_{\mu\nu})^2 ~+~
	\frac{1}{2}
		\Bigr(
			m\cdot [ \diff\wedge B]  ~-~
                        i\;\dual{F^a n^a}\Bigr)^2
 + i\,j_m B + i\,j^a_e A^a \, .
\eeq

Now, a few remarks concerning the Lagrangian (\ref{dualglue}) are in order:

(a) First, if the magnetic current is vanishing, $j_m=0$ then
the integration over the field $B$ reproduces
the standard Lagrangian of the gluodynamics
(we are omitting the gauge-fixing and ghost terms).

(b) As far as the quantization is concerned, the Lagrangian
(\ref{dualglue}) reproduces the correct degrees of freedom of the free
gluons. Indeed, in the limit $g\to 0$ and for $n^a=\delta^{a,3}$ the
Lagrangian (\ref{dualglue}) becomes:
$$
L_{dual}(A^a,B) ~=~ \frac{1}{4}(\diff\wedge A^1)^2 + \frac{1}{4}(\diff\wedge A^2)^2 +
i\,j_m B + i\,j^a_e A^a +
$$
\beq
+ \frac{1}{2}[m\cdot(\diff\wedge A^3)]^2 + \frac{1}{2}[m\cdot(\diff\wedge B)]^2 +
\eeq
$$
+\frac{i}{2} [m\cdot(\diff\wedge A^3)] [m\cdot\dual{(\diff\wedge B)}] -
\frac{i}{2} [m\cdot\dual{(\diff\wedge A^3)}] [m\cdot(\diff\wedge B)]\,,
$$
which is essentially the Zwanziger Lagrangian (\ref{Zw-action})
plus non-interacting non-diagonal gluons. Quantization at this point
is the same as in the case of a single photon.

(c) The emergence of the vector $n^a$
in the Lagrangian (\ref{dualglue}) is of crucial importance. The point is
that the origin of the vector $n^a$ goes back to choosing the color
orientation of the monopoles.  As is emphasized above the monopole solutions
are Abelian in nature which means, in particular, that they can be rotated
to any direction in the color space by gauge transformations.  Thus, picking
up a particular $n^a$ is nothing else but using the gauge fixing freedom.
Therefore, we can either average over the directions of $n^a$ or fix $n^a$
but evaluate only gauge invariant quantities, like the Wilson loop.

(d) The $Z_2$ nature of the monopoles is manifested in the freedom of
changing $n^a\to -n^a$, $B_\mu \to -B_\mu$. Indeed, under such
transformation a monopole with the charge $Q_m=+1$ is transformed into a
monopole with $Q_m=-1$ and vice versa. In the language we used above such a
transformation corresponds to adding a Dirac string with a double magnetic
flux.  We see that the averaging over $\pm n^a$ is a part of the overall
averaging over all possible embedding of the $U(1)$ into the $SU(2)$ gauge
group.

(e) In its generality, the Lagrangian (\ref{dualglue})
obviously possesses $SU(2) \times U(1)$ gauge invariance:
\beq
\label{freedom}
SU(2)\,:\;
	\begin{array}{ccc}
		F_{\mu\nu} & \to &  \Omega^{-1}  F_{\mu\nu}  \Omega  \\
		n^a \sigma^a \,=\, n & \to & \Omega^{-1}  n  \Omega  \rule{0mm}{5mm} \\
%		\chi^a \sigma^a \,=\, \chi & \to & \Omega^{-1}  \chi  \Omega \rule{0mm}{5mm} \\
	\end{array}\,,
\qquad
U(1)\,:\;
	\begin{array}{c}
		B_\mu  ~\to ~ B_\mu + \diff_\mu \alpha \\
	\end{array}\,.
\eeq

All the features (a)-(e) indicate that the Lagrangian (\ref{dualglue})
is a reasonable choice to describe the dual gluodynamics.  Now, we
would like to put the construction (\ref{dualglue}) on further tests.
An obvious check is the evaluation of the quantum corrections to the
heavy monopole potential $V_{\bar{m}m}$ at short distances. Indeed, it
can be shown, \cite{main}, that the radiative corrections result in the
standard running of the non-Abelian constant $g^2$.

\subsection{The 't~Hooft Loop and the Dual Gluon}
\label{nabmonrenorm}

We have already derived a continuum analog of the 't~Hooft loop, see
Eq(\ref{tHooft-general}).  Now, using the notion of the dual gluon field we
will turn to a more detailed analysis of the heavy monopole potential.

For a rectangular contour $T\times R$ we can write for the surface
$\Sigma^a$:  \beq \label{Sig-col-1}
\Sigma^a_j ~=~ n^a \frac{1}{(m\diff)} [m \wedge j]
\eeq
which is a particular solution to $\partial\Sigma=j$. It is obvious that the
current $j$ will serve now as a source of the dual gluon $B$.  Upon
substitution of (\ref{Sig-col-1}) the expectation value $H(j)$ is
represented as:

\beq
\label{H-2}
H(j) ~=~ \frac{1}{Z}\int \cD A \int\limits_{n^2=1} \!\!\! \cD n \;
\exp\{\; - \frac{1}{4 g^2} \int \mathrm{d}^4x \, (
F^a + 2\pi \,n^a\, \frac{1}{(m\diff)} \dual{[m\wedge j ]} )^2 \;\}\,,
\eeq

Note that Eq. (\ref{Sig-col-1}) extends the definition of the field $n^a$
into the entire space--time while only its value of $n^a$ on the string
$\Sigma_j$ does matter, while a particular way of extending $n^a(\sigma)
\to n^a(x)$ is irrelevant.  Moreover, the path integral in (\ref{H-2}) is to
be performed  with a constraint

\beq
\label{constr}
\varepsilon^{abc} n^b ( F^c \cdot \frac{1}{(m\diff)} \dual{[m\wedge j]})~=~0 , \qquad a=1,2,3\,,
\eeq
which is a consequence of the constraint~(\ref{tn}).
The constraint equation (\ref{constr}) may be implemented by an additional field $\chi^a$:
\beqn
\label{H-3}
H(j)= \frac{1}{Z}\int\limits_{n^2=1} \!\!\! \cD n \cD A \cD \chi \,
\exp\{\; - \frac{1}{4 g^2} \int \! \mathrm{d}^4x \,
(F^a +  \\
+ 2\pi \,n^a\, \frac{1}{(m\diff)} \dual{[m\wedge j ]} )^2
+ i\frac{2\pi}{g}\!\!\int\! j C \;\}\,, \nonumber
\eeqn
\beq
\label{C}
C_\mu ~=~ C_\mu(\chi, n, F )~=~
\frac{1}{(m\diff)} \varepsilon^{abc} \chi^a n^b ( m \cdot \dual{F}^c )_\mu\,.
\eeq
It is worth emphasizing that the field $n^a(x)$ is a kind of a fake variable
in the representation (\ref{H-3}). The path integral is clearly independent
on $n^a$ away from the string (\ref{Sig-col-1}), but at the same time
$n^a(x)$ for $x\in \Sigma_j$ is determined through (\ref{constr}).

Coming back to the evaluation of the potential $V_{\bar{m}m}$, on the
classical level we tend the coupling $g\to 0$. In this limit and for
$n^a=\delta^{a,3}$ the Lagrangian (\ref{dualglue}) reduces to:

\beq
\label{dual-L0}
L^0 ~=~ \frac{1}{4}[\diff\wedge A^a]^2 ~+~
\frac{1}{2}(m\cdot [ \diff\wedge B  ~-~ i\;\dual{\diff\wedge A^3}])^2\,,
\eeq
which essentially coincides with that of Zwanziger (\ref{Zw-action}).  In
particular, to the lowest order in $g^2$ the monopole-antimonopole
interaction is given by a single $B$-field exchange. Using
Eq.~(\ref{propagators}) one immediately obtains:

\beq
H(j) ~=~ \exp\{\;  - \frac{1}{2} \, (\frac{2\pi}{g} )^2 \, \int j_\mu \Delta^{-1} j_\mu \;\} ~\sim ~
\exp\{\;  T \frac{4\pi^2}{g^2} \Delta^{-1}_{(3)}(R)\;\}
\eeq
\beq
\label{tree}
V_{m\bar{m}}(R) ~=~ -\frac{1}{g^2}\cdot\frac{\pi}{R}
\eeq
Which is the expected result of course.

Having full Lagrangian (\ref{dualglue}) we can consider the radiative
corrections to (\ref{tree}). The calculations similar to that of Section
\ref{monoprenorm} give the result~\cite{main}:

\beq
\label{small-R-potential}
V_{m\bar{m}}(R) ~=~ - {\pi\over g^2(R)} \;{1\over R}\, ,
\eeq
which is valid of course at small distances. Eq. (\ref{small-R-potential})
makes manifest that monopoles in gluodynamics unify Abelian and non-Abelian
features. Namely, the overall coefficient, $\pi/g^2$ is the same as in
Abelian theory, while the running of the coupling $g^2(R)$ is the same as in
the non-Abelian theory. The $U(1)$ normalization is based on the
classification of the monopoles in non-Abelian theory (see, e.g.,
\cite{coleman} and references therein). The running of the coupling reflects
the general rule that the effect of the fluctuations at short distances can
be absorbed into the renormalization of the coupling. Eq.
(\ref{small-R-potential}) also show that the Dirac quantization condition is
valid at the perturbative level in non-Abelian theory.

In order to derive \eq{small-R-potential} we have to neglect the
contribution of the Dirac strings. We should subtract the
contribution of the strings, even if it is not vanishing within perturbation
theory. Essentially, the rule is that the product ${\bf H_1\cdot H}_2$
should not include the piece ${\bf H}^{Coul}\cdot {\bf H}^{Str}$.

To justify this rule we have to go actually beyond the Lagrangian
\eq{dualglue} and consider a regularization procedure which takes into
account the Dirac strings. It is worth emphasizing that within the lattice
regularization the term ${\bf H}^{Coul}\cdot {\bf H}^{Str}$ is indeed
absent. Let us recall the reader that in the lattice regularization the
Dirac string pierces negative plaquettes. This is true in the limit $g^2\to
0$, or $a\to 0$. If one looks for small deviation from $U_p=-1$, then the
action does not contain terms, which are linear in perturbations. Note that
this will be not true for the expansion around arbitrary $U_p \ne \pm 1$.
This result means in turn that in the continuum limit there is no term ${\bf
H}^{Coul}\cdot {\bf H}^{Str}$. Thus, our naive removal of the effects of
the Dirac string from the radiative corrections, is justified by the lattice
regularization.

The prediction of the monopole potential at large distances is also
possible~\cite{main} if we use the Abelian dominance model. The existing
numerical data~\cite{hoelbling} do not contradict the predictions of
Ref.~\cite{main}. Higher statistics is needed, however, to crucially check
the theory.

\subsection{Phenomenological Lagrangians}
\label{PhenLagr}

We now switch to the physics of large distances, or the dual superconductor
model of the confinement, see the Introduction.

Here, the basic lesson brought about by the lattice simulations is that the
monopoles condense (see reviews \cite{reviews} and Section~\ref{Lattice}).
In Subsection~\ref{latticereg} we have sketched theoretical
arguments~\cite{Polyakov-compact-U1} for the monopole condensation within
the compact $U(1)$. In case of the non-Abelian theory, a simple-minded
generalization of these arguments is as follows. Let us begin with a lattice
formulation and a small coupling $g^2_{SU(2)}$. Let us furthermore
integrate, a la Wilson, over small-scale fluctuations and go in this way to
a larger-size lattice. The only effect of this is the rescaling of the
coupling $g^2_{SU(2)}$ according to the rules of the renorm-group. However,
we could expect that this procedure would not work further once we reach
$g^2_{SU(2)}\sim 1$. Indeed, the same non-Abelian coupling $g^2_{SU(2)}$
governs the physics associated with any $U(1)$ subgroup. And we know that
for $g^2_{U(1)}\sim 1$ there is a phase transition due to the monopole
condensation. Thus, we expect the monopole condensation in non-Abelian
theory as well, unless something else happens at smaller $g^2_{SU(2)}$.

In the continuum limit and in the field theoretical language,
one describes usually the monopole condensation within a (dual)
Abelian Higgs model (AHM), see, e.g. reviews~\cite{baker,brambilla,reviews} and references therein.
The action of the model looks as:
\beq\label{AHM_action}
S= \int d^4x \left\{
\frac{1}{4Q_m^2} F^2_{\mu\nu} + \frac{1}{2} |(\diff - i B)\Phi|^2 +
\frac{1}{4} \lambda (|\Phi|^2-\eta^2)^2
\right\}
\eeq
where $Q_m$ is the magnetic charge, $B_{\mu}$ is an Abelian gauge field,
$F_{\mu\nu}$ is the corresponding field-strength tensor,
$F_{\mu\nu}\equiv\diff_{\mu}B_{\nu}-\diff_{\nu}B_{\mu}$, and $\lambda,\eta$ are constants.
The magnetically charged scalar field $\Phi$
condenses in the vacuum, $<\Phi> =\eta$, and the physical
vector and scalar particles are massive, $m^2_V=Q_m^2\eta^2, m_H^2= 2
\lambda \eta^2$.

The corresponding equations of motion possess an Abrikosov-type string solution.
An open string with electrically charged particles at the end points is
thought to mimic a confined $\bar{Q}Q$ pair in QCD. The string tension is
a function of the parameters of the theory, that is $m_V,m_H$.

There is one subtle point about the dual superconductor model which we would
like to emphasize. Namely, if we turn back to the simple arguments of the
Subsection~\ref{latticereg} which explain the monopole condensation, we can
easily visualize that, indeed, the monopole condensate is non-vanishing,
that is $<\Phi> =\eta\neq 0$. Moreover, it is then consequence of the gauge
invariant interaction that $m_V^2\neq 0$ as well. However, it is not easy to
understand at all which parameter is related to the Higgs mass, $m_H$. One
allows usually for arbitrary $m_V, m_H$ which are treated as fit parameters
to reproduce the lattice data. Moreover, the most recent fits to the
transverse structure of the confining string are close to the Bogomolny
limit, $m_V=m_H$ see, e.g.,~\cite{MIP} and Subsection~\ref{tube}. Of course,
there is an intrinsic uncertainty in all the fitting procedure since one
using the classical solutions to (\ref{AHM_action}) to fit the results of
the lattice simulations which account fully for the quantum effects.

From purely theoretical point of view another fitting procedure would be
more logical. Namely, at large distances it is logical to assume the London
limit since the monopole condensation in the compact $U(1)$ does correspond
to this limit. As for the finite Higgs mass $m_H$ it should be produced by
account for the non-Abelian gluons which are to be manifested at short
distances. However, there is no theoretical papers which would produce an
explicit procedure of this type and we cannot include any detailed
discussion of such an approach into this review. Although, as we shall see
in the next Subsection, further exploration of this approach seems urgent.

\subsection{Casimir Scaling}
\label{Casimir}

As we already mentioned in the Introduction, a direct use of the Abelian
models (\ref{AHM_action}) is in contradiction with the Casimir scaling.
Indeed, whatever $U(1)$ subgroup we choose there are quarks neutral with
respect to this subgroup once the (color) isospin of the quarks is integer.
However, after the experience with evaluation of the heavy monopole
potential (see above) this puzzle does not look very difficult to solve.
Namely, we can use a $U(1)$ language for dual gluons. However, we should
understand the choice of a particular $U(1)$ as a gauge fixing procedure and
evaluate gauge-invariant quantities.

Thus, combining the idea of introducing an effective Higgs field $\Phi$
with fundamental Lagrangian (\ref{dualglue}) we come
to the following effective theory of $SU(2)$ gluodynamics:
\beq
\label{Z-eff}
Z~=~\int \cD A \cD B \cD \Phi \;  \cD \chi \! \int\limits_{n^2=1} \!\!\! \cD n \;
\exp\{\; -\int \mathrm{d}^4x L_{\mathrm{eff}} \;\}\,,
\eeq
\beqn
\label{L-eff}
\displaystyle{ L_{\mathrm{eff}} = } &
        \displaystyle{\frac{1}{4}( F^a_{\mu\nu})^2 +
	\frac{1}{2} \Bigr[
		\Bigr(
                        m\cdot [ \diff\wedge B  -
                        i\;\dual{F^a n^a} -
			\varepsilon^{abc} \chi^a n^b \dual{F}^c \, ]
		\Bigr)_\mu\,
        \Bigr]^2 +} \\
%\rule{0mm}{7mm}                      &
        & \displaystyle{ + \frac{1}{2} |(\diff + i\frac{4\pi}{g} B)\Phi|^2 +
        V(|\Phi|)}\,, \nonumber
\eeqn
where
\beq
V(|\Phi|) ~=~ \lambda \left( |\Phi|^2 - \eta^2 \right)^2\,,
\eeq
and $\lambda$ and $\eta$ are phenomenological constants. Of course, the vacuum expectation value
of the Higgs, or monopole field is of order $\Lambda_{QCD}$. The physical assumption behind (\ref{Z-eff})
is that the effective
size of the monopoles with $|Q_m|=2$ is in fact numerically small,
although generically it is of order $\Lambda_{QCD}$ (see also~\cite{shuryak}).

So far we discussed only embedding of the $U(1)$ into $SU(2)$. As far as the
physics of large distances is concerned, there is a subtle point, how to
define the $U(1)$ dynamically. As is mentioned above the $U(1)$ arises in
lattice simulations through gauge fixing.  The monopole properties depend on
the particular choice of $U(1)$. The Maximal Abelian projection turns most
successful as far as the interaction of the heavy quarks in the fundamental
representation is concerned~\cite{reviews}. Thus, we will assume the use of
this gauge. Moreover, let us fix the orientation of the dual $U(1)$ group in
the color space as the rotations around the third axis. Then knowing the
classical, Abrikosov-type solution allows to find directly the string
tension, $\sigma_{Ab}( 2 T_3)$ as a function of the third component of the
isospin of the quarks, $T_3$ and of the parameters of the
Lagrangian(\ref{Z-eff}).  The role of the Wilson loop is to ensure that the
$Q\bar{Q}$ system is in the state with the total isospin $T=0$.  As a
result, the string tension measured via the Wilson loop is given by:

\beq\label{QM}
\sigma_T~=~{1\over 2T+1}\,\sum\limits_{T_3 = -T}^{T} \sigma_{Ab}( 2 T_3)\,,
\eeq
where $\sigma_{Ab}$ denotes the Abrikosov string tension, calculated in
pure Abelian Higgs model with and for external charges of the magnitude $\pm
2 T_3$ while the overall factor $(2T+1)^{-1}$ is due to the normalization of
the wave function.  Note that quantum mechanically Eq (\ref{QM}) is similar,
say, to expressing the hyperfine splitting for a state with a given total
spin in terms of the spin-spin interaction of the constituents.

In the London limit, $m_H\gg m_V$, the string tension $\sigma_{Ab}(2 T_3) \sim T_3^2$ and we
reproduce the Casimir scaling:
$$
\sigma_T~=~(const){1\over 2T+1}\,\sum\limits_{T_3 = -T}^{T} T_3^2 ~=~ (const)\frac{1}{3} \, T(T+1) \,.
$$
In the Bogomolny limit, $m_H=m_V$, $\sigma_{Ab}(2 T_3) \sim |T_3|$ and $\sigma_T\sim T(T+1)/(2T+1)$.

As is mentioned above, the structure of the confining string for the quarks
in the fundamental representation is best described in the Bogomolny limit,
see Subsection~\ref{tube} and Ref.~\cite{MIP}. As is mentioned in the
preceding Subsection, a possible way out of the difficulty could be account
for the non-Abelian gluons at short distances. Indeed, the string tension is
sensitive to much larger distances than the string structure in the
transverse direction.  At large distances, where the picture of the monopole
condensation applies, one expects the London limit to hold (see the
preceding Subsection).  At smaller distances the charged gluons could become
important.

We conclude this subsection with a remark, that from the theoretical point
of view, the most important limitation in the use of (\ref{L-eff}) is that
it can be consistently treated on the classical level only. The difficulty
to extend it to the quantum level is due to the Dirac strings. Indeed,
perturbatively the Dirac veto is violated for virtual particles (see
Subsections~\ref{monoprenorm}, \ref{nabmonrenorm}).  When the monopoles
condense the Dirac strings are filling the whole of the vacuum and there are
no known ways to rectify the perturbation theory.  In case of the Abelian
Higgs model, it is even more convenient to use the dual language when the
Dirac strings are attached to the electric charges.  Then, to respect the
Dirac veto one should impose the condition that the Higgs field vanishes
along a line connecting charges~\cite{ss}.  The static
Abrikosov-Nielsen-Olesen string satisfies this constraint~\cite{abrikosov}.
However, there is no known complete set of solutions satisfying this
boundary condition.

\subsection{Conclusions \# 3}

The use of the Zwanziger Lagrangian is a convenient means to describe
interaction of photon with both magnetic and electric charges. Formally, one
introduces two potentials, $A_{\mu}, B_{\mu}$. However, because of the
condition $F_{\mu\nu}(A)~=~\dual{F}_{\mu\nu}(B)$ there is a single photon in
fact. One can get rid of one of the potentials. But then the Dirac string
emerges explicitly and the formulation becomes non-local.

Another point, very important from the point of view of the phenomenological
applications, is that in case of the gluodynamics the dual gluon (that is
the field $B$) is still Abelian like although the ``direct'' gluons (that is
the fields $A^a$) are in an adjoint representation of the non-Abelian group.

An important conclusion concerning the phenomenological Lagrangians is that
observation of the Casimir scaling imposes the London limit, as far as one
fits the lattice data with classical solutions. To avoid contradictions with
fitting the structure of the confining string one should assume then that
the applicability of the AHM is limited at short distances by existence of
the non-Abelian gluons, while the parameterization in terms of a finite
Higgs mass, $m_H$ can be misleading. However, at present time there exists
no explicit realization of this idea.

\section{Monopoles in Lattice Gluodynamics} \label{Lattice}

In this section we give the brief review of the numerical results obtained
for monopoles on the lattice. We pay the special attention to two new
topics: structure of the confining string and structure of monopoles in the
Abelian projection. Most of the results are obtained for $SU(2)$ lattice
gauge theory for the maximal Abelian projection. The main definitions are
given in Section~\ref{DefLatMon}.

\subsection{Properties of Abelian Monopoles in $SU(2)$ Lattice
Gluodynamics}

At first we list the main properties of the Abelian monopoles in $SU(2)$
lattice gluodynamics in the maximal Abelian projection~\cite{MAAP},
the reader can find many important details in reviews~\cite{reviews}.

\begin{figure}[htb]
\vskip1.7cm
\begin{center}
\begin{tabular}{cc}
\hskip-3mm{\epsfxsize=0.4\textwidth\epsfbox{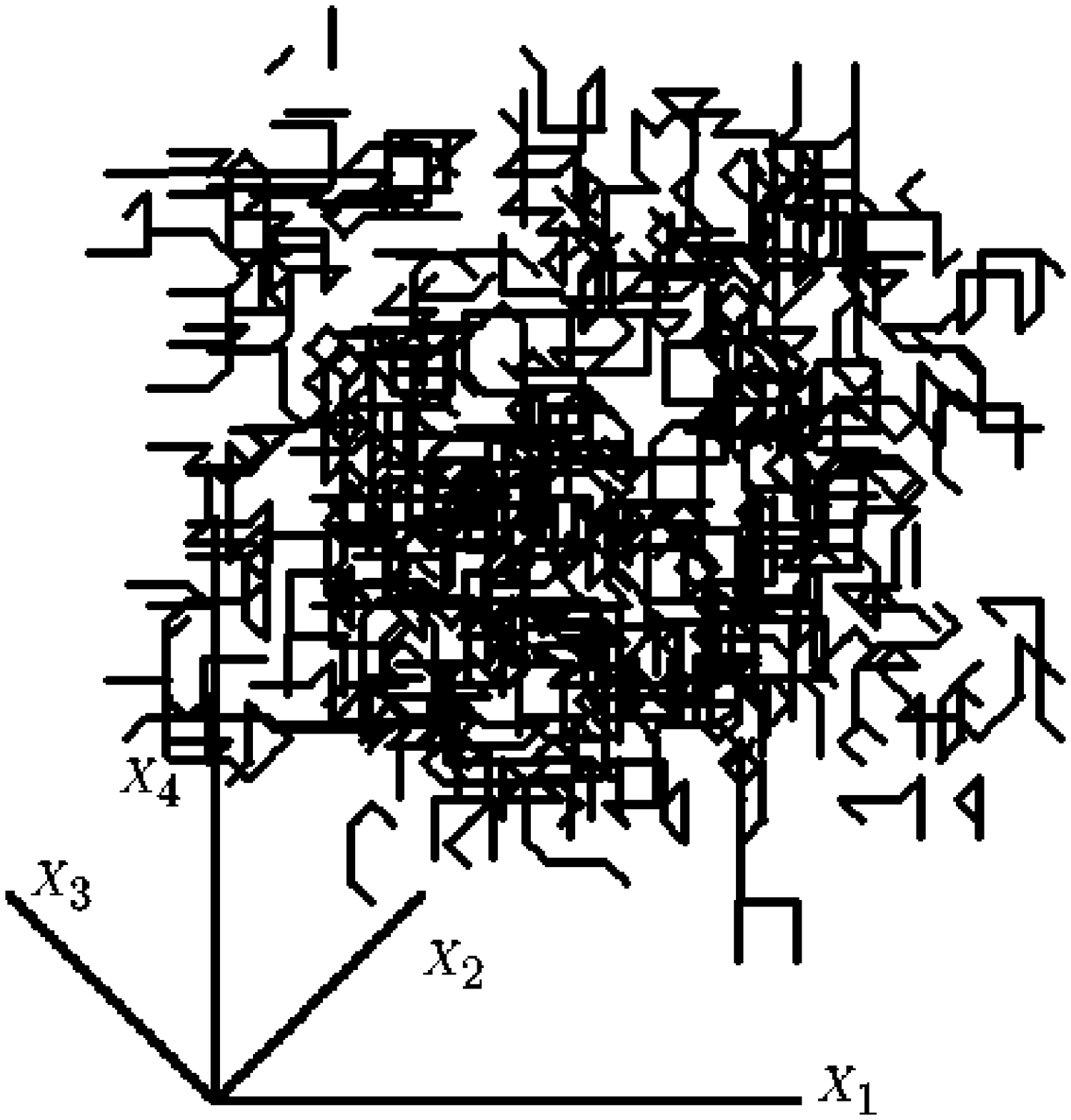}}&\hskip8mm
{\epsfxsize=0.4\textwidth\epsfbox{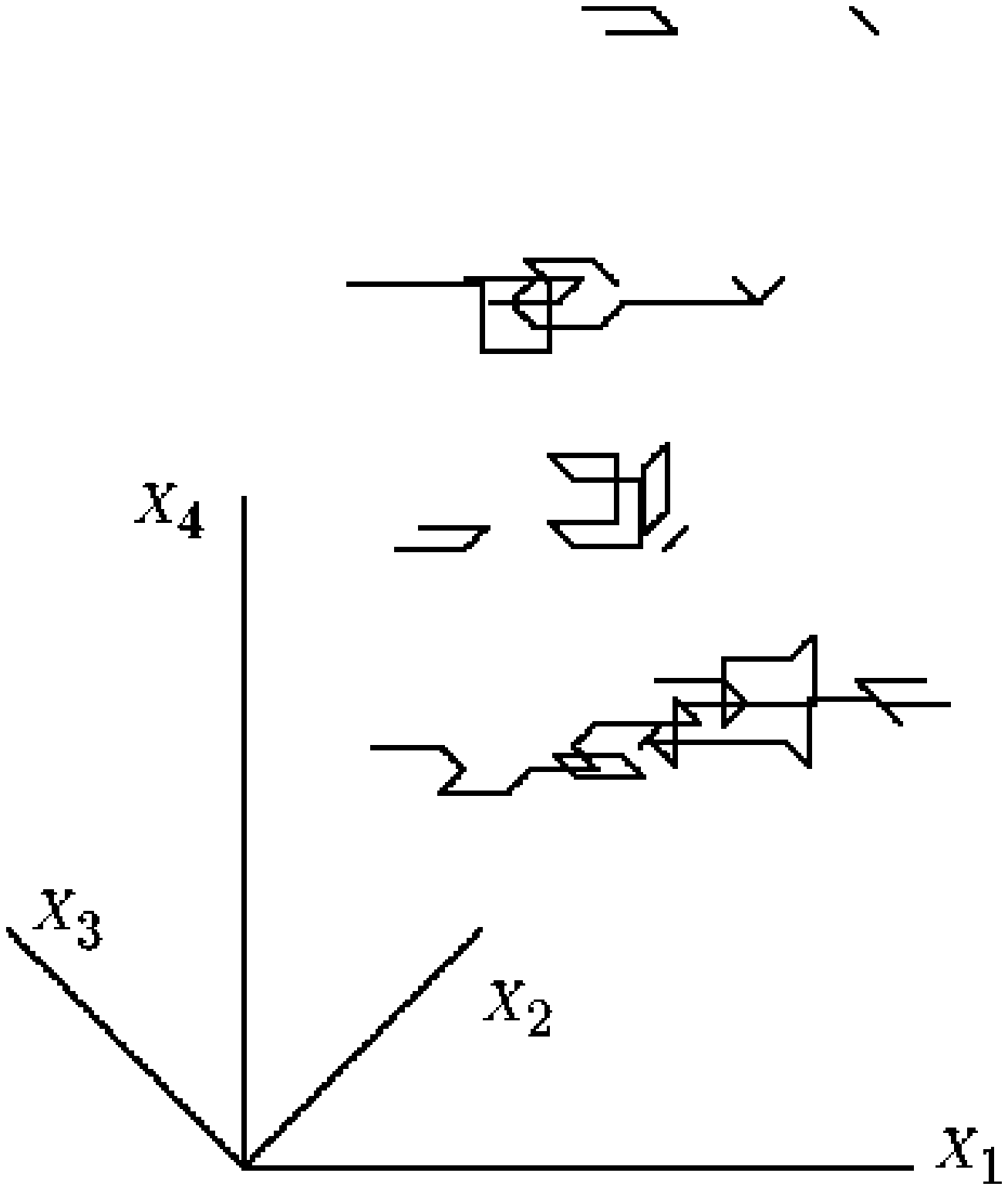}}\\
 (a) & (b)\\
\end{tabular}
\end{center}
\vskip-5mm
\caption{The abelian monopole currents for the confinement (a)
($\beta=2.4$, $10^4$ lattice) and the deconfinement (b) phases
($\beta=2.8$, $12^3 \cdot 4$ lattice). All monopole currents are closed
(conserved) due to the periodic boundary conditions.} \label{CurrentsConf}
\end{figure}

{\em 1.}
In the confinement phase the monopole currents form a dense cluster,
we call it infrared (IR) cluster, and there is a number of small mutually
disjoint clusters, "ultraviolet" (UV) clusters. In the deconfinement phase
the monopole currents are dilute. In Figure~\ref{CurrentsConf} we
demonstrate these facts showing the abelian monopole currents for the
confinement (a) and the deconfinement (b) phases. The IR cluster
percolates~\cite{percolation} and has a nontrivial fractal
dimension~\cite{fractalD}, $D_f >1$. The properties of UV clusters differs
much from those of the IR cluster, it can be shown that the IR monopole
cluster is responsible for the confinement of quarks~\cite{clusters}.

{\em 2.} The $SU(2)$ string tension is well reproduced by the
contribution of the abelian fields and/or abelian monopole
currents~\cite{Abelian-Dominance}. This property is called monopole
(Abelian) dominance. While the Abelian dominance is an almost trivial
property of the system~\cite{abdomintr}, the monopole dominance indicates
that Abelian monopoles are important degrees of freedom for the infrared
physics.

{\em 3.}
The monopole condensate is nonzero in the confinement phase of $SU(2)$
gluodynamics and it vanishes at the critical temperature corresponding
to the deconfinement phase transition, thus it plays the role of the
order parameter~\cite{OrderP}. We can treat this fact as a
justification of the dual superconductor model of the QCD vacuum, the
monopoles playing the (dual) role of the Cooper pairs.

{\em 4.} The monopoles are correlated with the density of $SU(2)$ action.
The total action of $SU(2)$ fields is correlated with the total length of
the monopole currents~\cite{ShSu95}, so there exists a global correlation.
The Abelian monopoles in the MaA projection are also locally correlated with
the non-Abelian action density~\cite{physobj}. This fact shows that
monopoles are some physical objects (not the artifacts of the singular gauge
transformation), since by definition we call the object physical if it
carries the action.

{\em 5.} The correlations of the monopole currents,
the electric currents, the topological charge density and the action
density was found in Refs.~\cite{monopcorr}.

{\em 6.} In Refs.~\cite{monLagr} the effective Lagrangian for monopoles
was reconstructed from numerical data for monopole currents for
$SU(2)$ gluodynamics in the Maximal Abelian gauge. It occurs that this
Lagrangian corresponds to the Abelian Higgs model, the monopoles are
condensed and the {\it classical} string tension of the
Abrikosov-Nielsen-Olesen string describes well the {\it quantum}
string tension of $SU(2)$ gluodynamics. It means that the monopole
degrees of freedom are important for the description of the
gluodynamics at large distances.

\subsection{Anatomy of the Confining String} \label{tube}

\begin{figure}
\epsfxsize=7.5cm
\centerline{\epsfbox{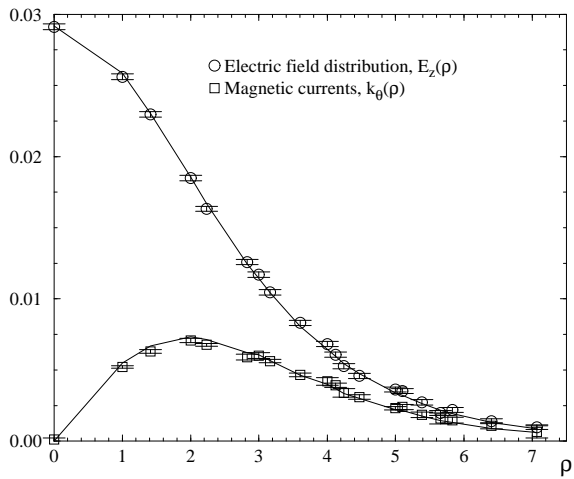}}
\caption{The fit of the lattice data
by the classical
solution of the discretized version of the Abelian Higgs model
{\eq{AHM_action}}.  \label{balifit}}
\end{figure}

The authors of Ref.~\cite{stringstruct} presented the results of the
numerical study of the confining string in $SU(2)$ lattice gluodynamics.
They measured the expectation value of the electric field, $\vec{E}$, and
the expectation value of the monopole current, $\vec{k}$, near the line
connecting the test quark--antiquark pair. For the test quarks placed at the
$z$ axis, only the $z$ component, $E_z(\rho)$, of $\vec{E}$ and the
azimuthal component, $k_\theta(\rho)$, of $\vec{k}$ were found to be
non-vanishing (here $\rho=\sqrt{x^2+y^2}$ is the distance from the center of
the flux tube and the azimuthal angle $\theta$ is defined as usual: $\tan
\theta=y/x$). It occurs\cite{MIP} that these data can be perfectly described
by the solution of the {\em classical} equations of motion for the Abelian
Higgs model \eq{AHM_action}. In Fig.~\ref{balifit} by solid lines we show
$E_z$ and $k_\theta$ for the Abrikosov string of the Abelian Higgs model
\eq{AHM_action}. The discretized version of the classical equations of
motion is used in order to imitate the coarseness of the lattice used in
numerical simulations~\cite{stringstruct}.

The parameters
of the model corresponding to the best fit of the numerical data are:
\beqn\label{param-values-lattice}
Q_m/2\pi &  = &   0.9519 \pm  0.0041 \, , \\
m_V & = & 0.4522 \pm   0.0206 =  (1.0351 \pm 0.0472) \,{~GeV}\, , \\
m_H & = &   0.4747 \pm   0.0600 =  ( 1.0866 \pm  0.1373 ) \,{~GeV} \,\, .
\eeqn
It appears that vector and scalar masses are equal to each other within the
errors, thus we are at the so-called Bogomolny limit of the model.
The Abrikosov string tension turns out to be
\beq
\label{sigma-lattice-fit}
\sqrt{\sigma} \; =  \; 0.1808 \pm 0.0213 \; \approx
\; ( 414.8 \pm 48.9 ) \mbox{~MeV} \; \approx \; 0.94
\; \sqrt{\sigma_{SU(2)}} \, \, ,
\eeq
here $\sigma_{SU(2)}$ is the string tension of $SU(2)$ gluodynamics. To get
the dimensional quantities we have used $\sqrt{\sigma_{SU(2)}} = 440\, Mev$.

\subsection{Anatomy of $SU(2)$ Monopole}

At small values of the bare charge (at large values of $\beta$) the
compact electrodynamics is in the deconfinement phase, and
gluodynamics is in the confinement phase. On the other hand, at large
values of $\beta$ in the maximal Abelian projection $\cos\phi_l$ (we
use the parameterization \eq{Uparameters}) is close to unity (due to
\eq{MAAPdef} and \eq{Rphi}), and the gluodynamic plaquette action
$\frac 12 tr U_P$ is close to the action of cQED, $cos\theta_P$.  Why
the monopoles are not condensed in cQED and are condensed in
gluodynamics if the actions of both theories are close to each other?
It occurs that monopoles in gluodynamics have nontrivial structure and
even near the continuum limit they differ much from the Abelian
monopoles. The action of monopoles in gluodynamics is smaller than the
action of monopoles in cQED. In Refs.~\cite{compensation} it was found
that the action of the non-diagonal gluons, $S^{off}$, on the
plaquettes near the monopole is negative, and the full non Abelian
action, $S^{SU(2)}=S^{off}+S^{Abel}$, is smaller than the Abelian part
of the action. Thus the action of monopoles in gluodynamics is smaller
than that in Abelian theory and this is the possible explanation of
the fact why the Abelian monopoles in gluodynamics are condensed at
any value of the bare coupling.

As we discussed in Section \ref{zeroactmon} there exists the pure
gauge field \eq{zeroA}, \eq{OcrosO} which Abelian ($A^3_\mu$)
component contains Dirac monopole. The lattice analogue of this field
has zero non-Abelian action since the action of the Dirac string is
zero on the lattice. And this configuration presumably
corresponds~\cite{main} to the fields inside Abelian monopoles in
gluodynamics. At large distances due to quantum fluctuations the
fields around monopole may have a finite non-Abelian action density.

\begin{figure}
\epsfxsize=7.5cm
\centerline{\epsfbox{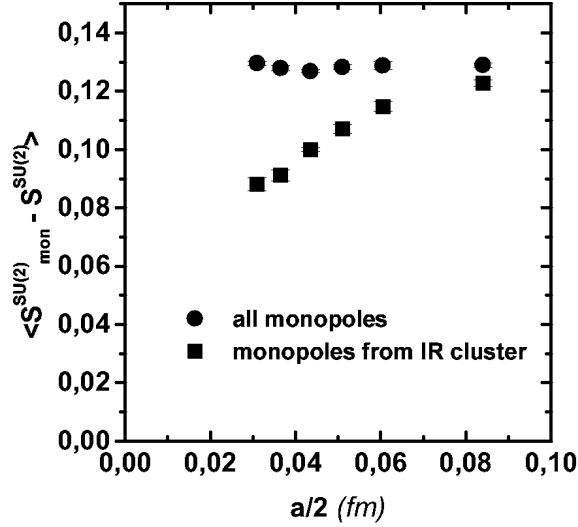}}
\caption{The dependence of $\bar{S}^{SU(2)}$ on $a/2$, the calculations are
performed on the lattices up to $30^4$.}
\label{nonabeact} \end{figure}

Below we present the recent results~\cite{monanat} of numerical
calculations in lattice $SU(2)$ gauge theory which confirm the above
discussion (see that paper for details).
In Fig.~\ref{nonabeact} we show the dependence of
$\bar{S}^{SU(2)} = <S^{SU(2)}_{mon} - S^{SU(2)}>$ on the half of the
lattice spacing, $a/2$, in fermi\footnote{We find the correspondence
between the bare charge and lattice spacing fixing the value of the
string tension $\sigma = 440\,\, \mbox{MeV}$ and using the numerical
data~\cite{su2scale} for the string tension in lattice units,
$\sigma\cdot a^2$.}. The explanation of the scale of the horizontal
axis is: $<S^{SU(2)}_{mon}>$ is measured on the plaquettes which are
faces of the cube dual to monopole, thus in a sense we are measuring
the average field strength on the distance $a/2$ from the monopole
center.

\begin{figure}
\epsfxsize=7.5cm
\centerline{\epsfbox{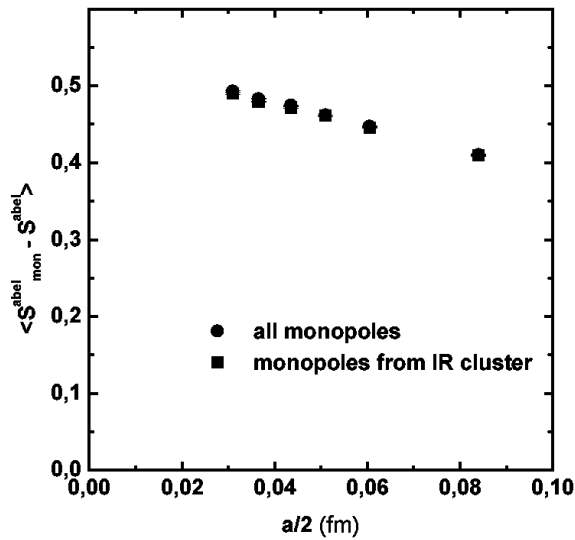}}
\caption{The same as in Fig.~\ref{nonabeact} for $\bar{S}^{Abel}$.}
\label{abeact}
\end{figure}

The circles on Fig.~\ref{nonabeact} correspond to the calculation which take
into account all monopoles, the squares correspond to monopoles taken from
the percolating cluster (IR cluster). The results of the analogous
calculation of the Abelian action near the monopole, $\bar{S}^{Abel} =
<~S^{Abel}_{mon} - S^{Abel}>$, are presented in Fig.~\ref{abeact}.

The results which follow from Figs.~\ref{nonabeact},~\ref{abeact} can be
summarized as follows:

\begin{itemize}

\item $\bar{S}^{SU(2)}$ at the center of the monopole belonging to the IR
cluster is compatible with zero (at least it decreases when we approach the
center). Thus at the center of monopole the fields are close to our
prediction, eq.~\eq{OcrosO}. The small value of $\bar{S}^{SU(2)}$ is the
reason why IR monopole clusters percolate. Note that the percolation
reflects the existence of the monopole condensate.

\item $\bar{S}^{Abel}$ for monopoles belonging to IR and UV
clusters is approximately the same. Thus there is no difference in the
Abelian part of the monopole fields in IR and UV clusters.

\item $\bar{S}^{Abel}$  increases when we approach the center of the
monopole. The closer we are to the center (the larger $\beta$) the larger
lattice we have to use since the properties of the IR monopole cluster are
strongly affected  by the finite volume effects~\cite{clusters}. This
technical difficulty prevents us to decide whether the Abelian action
density diverges or finite at the center of the monopole.

\end{itemize}

Note that these facts (and also Figs.~\ref{nonabeact},~\ref{abeact}) are
given for the action densities in lattice units, if some action density is
constant in lattice units, it corresponds to  $1/a^4$ (Coulombic) behavior
of this density in the continuum limit.

\subsection{Conclusions \# 4}

The results of numerical experiments in $SU(2)$ lattice gauge theory confirm
the validity of the dual superconductor model of the gluodynamic vacuum. The
monopoles are condensed in the confinement phase, and already at the
classical level are responsible for 94\% of the $SU(2)$ string tension. The
structure of monopoles in gluodynamics is nontrivial, the non-diagonal gluons
reduce the monopole action.

%======================================================================
\section*{Acknowledgments}

The authors are thankful to V.G.~Bornyakov, T.~Suzuki and A.I.~Veselov
for valuable discussions. M.N.Ch. and M.I.P. acknowledge the kind
hospitality of the staff of the Max-Planck Institut f\"ur Physik
(M\"unchen), where the work was initiated.  Work of M.N.C., F.V.G. and
M.I.P. was partially supported by grants RFBR 99-01230a, RFBR
01-02-17456, INTAS 96-370, JSPS Grant in Aid for Scientific Research
(B) (Grant No. 10440073 and No. 11695029) and CRDF award RP1-2103.

%========================================================================

\end{document}